\documentclass[useAMS,usenatbib,usegraphicx]{mn2e}
\DeclareGraphicsExtensions{.jpg, .eps}
\usepackage{fixltx2e}
\usepackage{subfig}
\usepackage{float}
\usepackage{upquote}

\title[Dynamics of stars around spiral arms]{The dynamics of stars around spiral arms}
\author[Grand et al.]
{Robert J.J. Grand $^1$\thanks{rjg2@mssl.ucl.ac.uk}, Daisuke Kawata $^1$ and Mark Cropper $^1$\\
$^1$ Mullard Space Science Laboratory, University College London, Holmbury St. Mary, Dorking, Surrey, RH5 6NT}

\pagerange{\pageref{firstpage}--\pageref{lastpage}} \pubyear{2011}

\begin{document}

\label{firstpage}
\maketitle

\begin{abstract}
Spiral density wave theory attempts to describe the spiral pattern in spiral galaxies in terms of a long-lived wave structure with a constant pattern speed in order to avoid the winding dilemma. The pattern is consequently a rigidly rotating, long-lived feature. We run N-body simulations of a giant disc galaxy consisting of a pure stellar disc and a static dark matter halo, and find that the spiral arms are transient features whose pattern speeds decrease with radius, in such a way that the pattern speed is almost equal to the rotation curve of the galaxy. We trace particle motion around the spiral arms. We show that particles from behind and in front of the spiral arm are drawn towards and join the arm. Particles move along the arm in the radial direction and we find a clear trend that they migrate toward the outer (inner) radii on the trailing (leading) side of the arm. Our simulations demonstrate that because the spiral arm feature is co-rotating, the particles continue to be accelerated (decelerated) by the spiral arm for long periods, which leads to strong and efficient migration, at all radii in the disc.
\end{abstract}


\section{Introduction}
Spiral density wave theory, first postulated by \citet{L60} and carried forward by \citet{LS64}, has been the most widely accepted theory of the spiral arm for almost 50 years. This theory considers the spiral arm to be the crest of a stellar density wave that rotates around the centre of the Galaxy at an angular pattern speed, $\Omega _p$, that does not vary with galacto-centric radius. Spiral density wave theory naturally explains the long-lived nature of the spiral arm. However, short-lived spiral arms were proposed by \cite{GLB65}, \cite{JT66} and \cite{T81}. These studies argue that spiral structure can grow from local instabilities and shearing rotation predicted from a flat rotation curve of disc stars that help to develop the spiral features \citep{A84}. Although these spiral arms are short-lived features that continuously disappear, they are regenerated by local growth of instability. In these transient spiral theories, the spiral structure itself can be self-excited in small over-dense regions as a consequence of shot noise \citep{T90}. If the over-density increases above a few percent, the growth becomes very non-linear and results in rapid growth of the spiral arm density (e.g. \citealt{T81}; \citealt{DT94}; \citealt{B03}; \citealt{Se10}). Non-linear growth of transient rigid waves is also demonstrated in the context of a shearing sheet \citep{F05}. While shearing is important for non-linear growth of spiral arms, it is also reported to be the main factor in the appearance of spurs and spiral arm bifurcation \citep{DB08}. 

More recently, the development of high-resolution numerical N-body/SPH simulations has played an important role in the study of spiral structure formation and evolution. Many numerical simulations including self-gravity and with no addition or forcing of spiral potential show short-lived transient spiral arms, which come and go throughout the evolution such that the spiral pattern of the disc is never removed (e.g. \citealt{SC84};  \citealt{BAM09}; \citealt{QDBM10}; \citealt{Fu11}; \citealt{WBS11}). \citet{SC84} note that the spiral pattern in N-body simulations generally fades over time because the spiral arm structure heats the disc kinematically and causes Toomre's Q parameter to rise, hence the disc becomes too stable against the development of non-axisymmetric structure. They suggest that continuous addition of a kinematically cold population of stars is crucial to maintain the spiral arm pattern. \citet{CF85} demonstrate this with gas clouds in circular orbit that are allowed to become future sites of star formation. Recently \citet{Fu11} suggest that the rapid disappearance of the pattern probably results from a low number of particles in these simulations, which leads to higher Poisson noise and more rapid growth of strong spirals, which then heat the disk and erase the spiral pattern. They demonstrated that the spiral pattern can be sustained without dissipation or continuous addition of cold components if a sufficiently high number of particles ($>10^6$) is used to prevent a large Poisson noise and therefore fast disc heating. Still, there are no numerical simulations that reproduce long-lived spiral arms like those predicted by spiral density wave theory \citep{Se11}, and this is the case for both spiral and barred-spiral galaxies (e.g. \citealt{BAM09}; \citealt{QDBM10}; \citealt{WBS11}). 

In spiral density wave theory, the angular pattern speed of the spiral arm is constant as a function of radius. The flat rotation curve of spiral galaxies implies the decreasing angular velocity of stars. Therefore, there is one radius where the stars rotate at the same velocity as the arm. This is called the co-rotation radius, inside which stars rotate faster than the spiral arm and outside which they rotate more slowly. The co-rotation radius marks a point of resonance, where stars very close to this radius have been shown to undergo large changes in their angular momentum and migrate radially to another part of the disc. This was first proposed by \citet{LBK72}, who showed that radial migration around co-rotation can occur without increasing the random motions of stars. \citet{SB02} extended this theory, and showed that transient arms provide a good mechanism for mixing, where each transient arm possesses a unique co-rotation radius around which nearby stars can migrate. The main conclusions of this paper were that only around the radius of co-rotation is there large angular momentum transport of individual stars without heating. There have been many studies on the effect of co-rotation resonances on structure, radial heating and mixing in spiral galaxies (e.g. \citealt{CS85}; \citealt{AL97}; \citealt{SB02}; \citealt{MQ06}; \citealt{R08}a; \citealt{RDS08}b; \citealt{SBl09}; \citealt{MFC11}) and in barred-spiral galaxies where a bar resonance is also included (e.g. \citealt{MF10}; \citealt{MA11}).
 

Very recently, there have been observational studies with evidence of a pattern speed that varies with radius, which is different from the constant pattern speed of spiral density wave theory. The evidence comes from a range of techniques and sources that include observations of no offset between different star formation tracers in nearby galaxies \citep{FR11}, and radially decreasing spiral arm pattern speeds deduced by the Tremaine-Weinberg method (e.g. \citealt{MRM06}; \citealt{MRM08}; \citealt{MRM09}; \citealt{SW11}), which inevitably leads to short-lived arms as a consequence. Numerical simulations of an interacting galaxy in \citet{DTP10} also shows a pattern speed that decreases as a function of radius. During the preparation of this work, \citet{WBS11} performed simulations of an isolated galaxy, and found a pattern speed that traces very well the rotation curve, which indicates that there is co-rotation of the spiral pattern with the stars at all radii. 

In this study, we analyse the pattern speed of the spiral arm feature and the dynamics of particles around the arm using our N-body simulations of a pure stellar disc. We determine the nature of the spiral arm pattern speed, and monitor how stellar motion is affected by the arm. We also find that the spiral arms in the simulations are transient and that the pattern speed is always similar to the circular velocity. We offer a qualitative insight into how phenomena such as radial migration might occur with the transient co-rotating spiral arms, and how the co-rotating spiral arm may evolve.

In Section 2 we explain how we set up the model and the initial parameters that we choose. In Section 3 we present the results of our analysis, compare them with previous studies and discuss their implications. In Section 4 we summarise the significance of the results and remark upon the value of the simulations and future work.

\section{Method and Model Setup}

Our simulations are performed with a Tree N-body code, GCD+ \citep{KG03}. We set up a disc galaxy that consists of a dark matter halo and a pure stellar disc with no bulge, and which is similar (slightly smaller) in size to the Milky Way. We describe the stellar disc component as a collisionless N-body system of particles, and adopt a static dark matter halo potential.

The dark matter halo density profile follows that of \citet{NFW97}:

\begin{equation}
\rho _{\rm dm} = \frac{3 H_0^2}{8 \pi G} (1+z_0)^3 \frac{\Omega _0}{\Omega (z)} \frac{\rho _{\rm c}}{ cx(1+cx)^2} ,
\end{equation} 
where $\rho _{\rm c}$ is the characteristic density described by \citet{NFW97}, the concentration parameter, $c =\rm \frac{r_{\rm 200}}{r_{\rm s}}$, and  $x= \frac{r}{r_{\rm 200}}$. The scale length is $r_{\rm s}$, and $r_{\rm 200}$ is the radius inside which the mean density of the Dark Matter sphere is equal to 200$\rho _{\rm crit}$:

\begin{equation}
r _{200} = 1.63 \times 10^{-2} ( \frac{M_{200}}{h^{-1} M_{\odot}} )^{\frac{1}{3}} [\frac{\Omega _0}{\Omega (z_0)}]^{\frac{-1}{3}} (1+z_0)^{-1} h^{-1} \rm kpc.
\end{equation} 
We assume $M_{200} = 1.7 \times 10^{12}$ $\rm M_{\odot}$, $c=15$, $\Omega _0 = 0.266$, $z_0=0$ and $H_0=71$ $\rm km$ $\rm s^{-1}$ $\rm Mpc^{-1}$. We do not apply adiabatic contraction for the dark matter halo, for simplicity.

\begin{figure}
\centering
\includegraphics[scale=0.43]{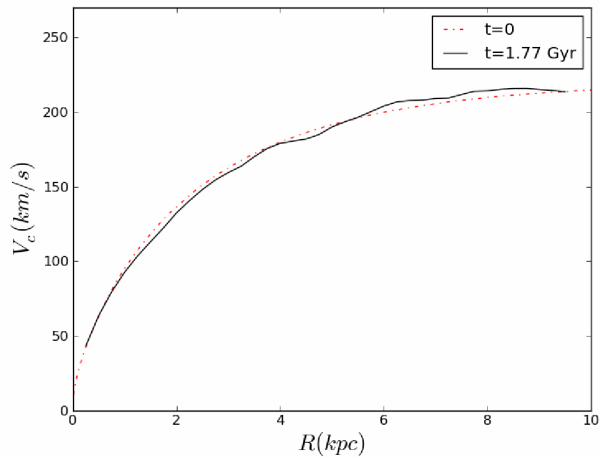}   
\caption{The initial circular velocity (dot-dashed red line) and the circular velocity at t$=1.77$ Gyr (solid black line).}
\label{vcrotc}
\end{figure}

\begin{figure}
\centering
\includegraphics[scale=0.43]{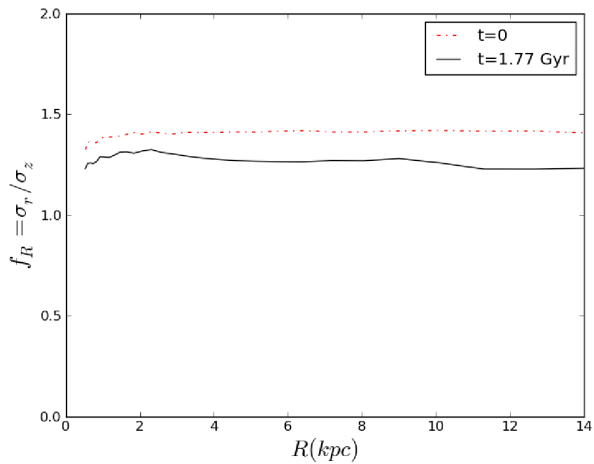}   
\caption{The ratio of velocity dispersions in the radial and $z$ direction, $f_R = \sigma_R / \sigma_z$, at t=0 (dot-dashed red line) and t = 1.77 Gyr (solid black line), plotted as a function of radius.}
\label{frQ}
\end{figure}

\begin{figure}
\centering
\includegraphics[scale=0.43]{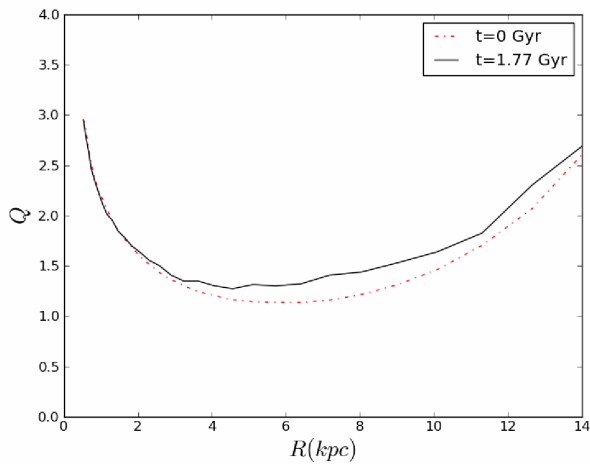}   
\caption{Toomre's instability parameter, Q, computed at t = 0 (dot-dashed red line) and t = 1.77 Gyr (solid black line), as a function of radius.}
\label{newQ2}
\end{figure}

The stellar disc is assumed to follow an exponential surface density profile:

\begin{equation}
\rho _{\rm d} = \frac{M_{\rm d}}{4 \pi z_d R_{\rm d}} {\rm sech}^2 (\frac{z}{z_d}) {\rm exp}(-\frac{R}{R_{\rm d}}).
\end{equation} 
We apply the mass of the disc, $M_d = 3 \times 10^{10}$ $\rm M_{\odot}$, the scale length, $R_{\rm d} = 3.0$ kpc and the scale height $z_d = 0.35$ kpc, which is constant over the disc. These parameters lead to a stellar surface density of $\Sigma=37$ $\rm M_{\odot}$ $\rm pc^{-2}$ at 8 kpc, which is similar to $\Sigma= 35.5$ $\rm M_{\odot}$ $\rm pc^{-2}$ obtained for disc stars in the solar neighbourhood \citep{FHP06}. 

The initial rotation curve is shown in Fig. \ref{vcrotc}. The rotation speed at 8 kpc is 210 $\rm km$ $\rm s^{-1}$, which is slightly lower than the generally accepted value of 220 $\rm km$ $\rm s^{-1}$ or even higher (e.g. \citealt{Mc11}) for the Milky Way. Although we constructed a galaxy similar in size to the Milky Way, it is not the intention of this study to reproduce the spiral arms of the Milky Way. We also deliberately choose a disc galaxy model that does not develop an obvious bar structure in order to avoid complexity resulting from the bar potential, hence we can focus more on the pure effect of spiral arm development. \citet{Fu11} show that if more than one million particles are used to describe the disc component, artificial heating that suppresses the spiral arm formation is not significant. We use $3 \times 10^{6}$ particles for the disc component, therefore the mass of each particle is $10^4$ $\rm M_{\odot}$. We adopt a fixed softening length of 340 pc, with the spline softening suggested by \citet{PM07}.

The velocity dispersion for each three dimensional position of the disc is computed following \citet{SMH05} to construct the almost equilibrium condition. One free parameter in this method is the ratio of the radial velocity dispersion to the vertical velocity dispersion, $f_R$, which relates as $f_R = \sigma _R / \sigma _z$. We choose $f_R=\sqrt{2}$ in the simulation shown. This is slightly lower than $f_R \sim 2$ that is the observed ratio for the Milky Way (e.g. \citealt{HNA09}; \citealt{Bi10}). However, as mentioned above, we do not aim to create a Milky Way-like galaxy. 

\section{Results and Discussion}

The simulation set up in Section 2 is evolved for 2 Gyr. As with the previous studies described in Section 1, we also find that the disc develops transient and recurrent spiral arms. In this paper, we focus on one arm identified around $t=1.77$ Gyr. The circular velocity at $t=1.77$ Gyr is shown in Fig. \ref{vcrotc}, which is not significantly different from the initial circular velocity. Fig. \ref{frQ} shows $f_R$ as a function of radius at t = 0 and t = 1.77 Gyr. The value drops slightly with time as the disc heats up slightly during evolution. This is quantified in Fig. \ref{newQ2}, which shows a slight growth of Toomre's instability parameter, Q, at the same time step. However, there is not significant heating, which helps to maintain spiral arms \citep{Fu11}.

Although we mainly discuss the analysis around $t=1.77$ Gyr in this paper, we also applied similar analyses to other spiral arms that developed at different times in this simulation as well as spiral arms in other simulations with different initial configurations of the disc and dark matter halo. We find that all the spiral arms we analysed show very similar results to those shown in this section. We also find similar results in several (barred and non-barred) simulations that take gas and star formation into account. These will be described in a forthcoming paper.  

First, we present the analysis and results of the pattern speeds of the chosen spiral arm. Then we examine the motion of selected particles around the arm, and present and discuss an analysis of their angular momentum and energy evolution. 


\begin{figure*}
 \centering \hspace{0.0mm}
 \subfloat{\includegraphics[scale=0.44]{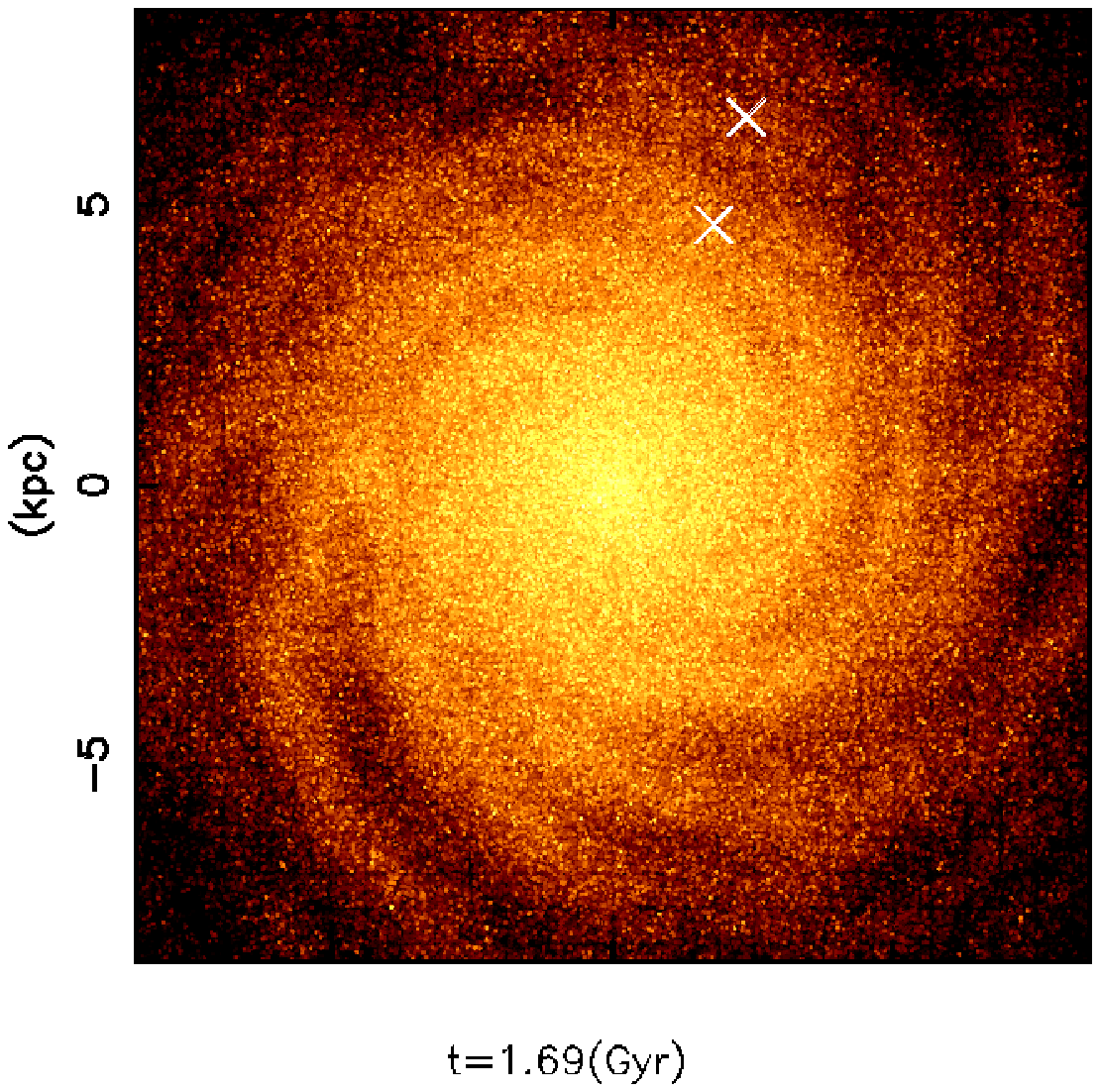}} \hspace{5.0mm}  
  \subfloat{\label{fig:t=1.75Gyr}\includegraphics[scale=0.44]{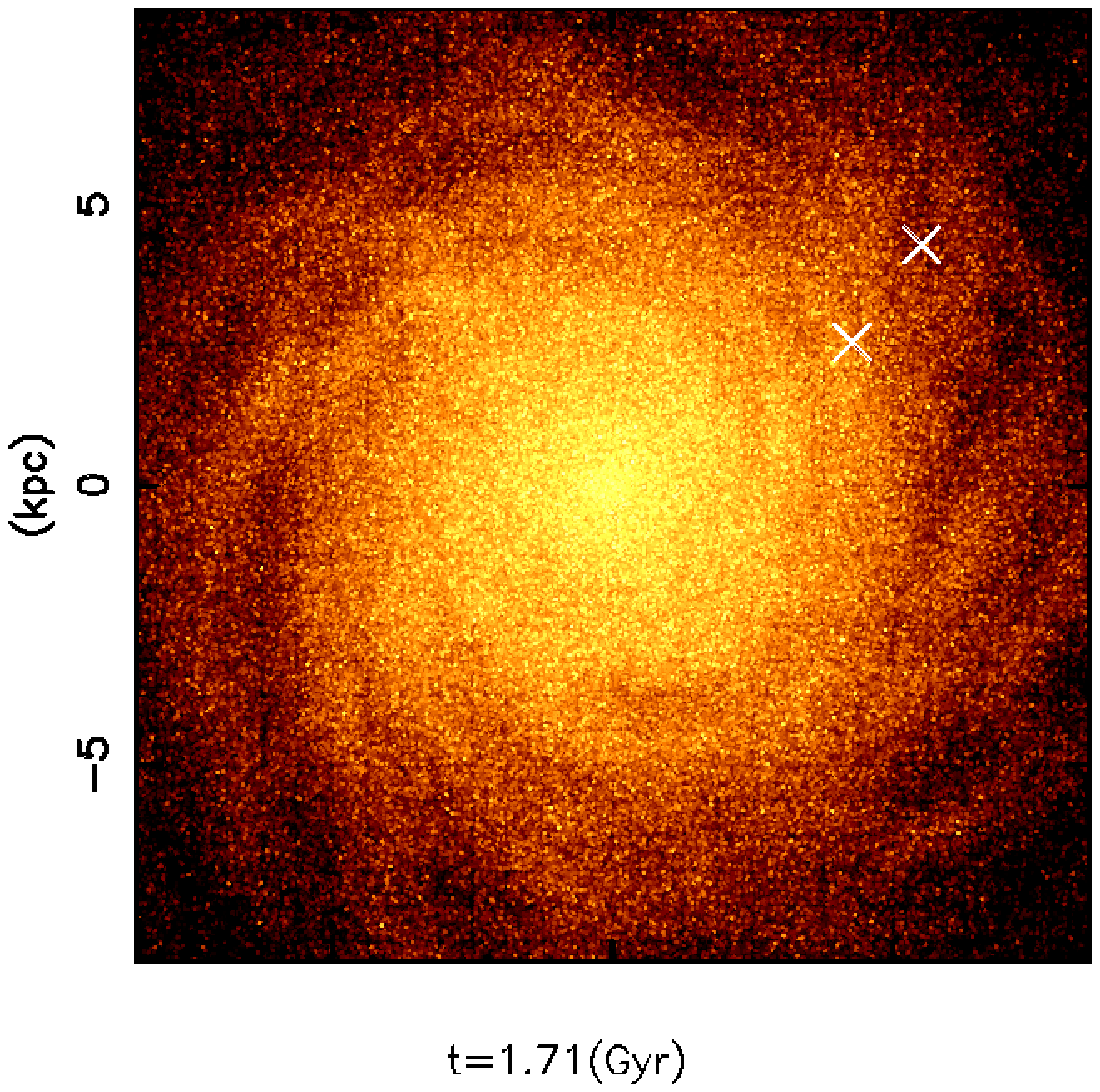}} \hspace{5.0mm} 
  \subfloat{\label{fig:t=1.77Gyr}\includegraphics[scale=0.44]{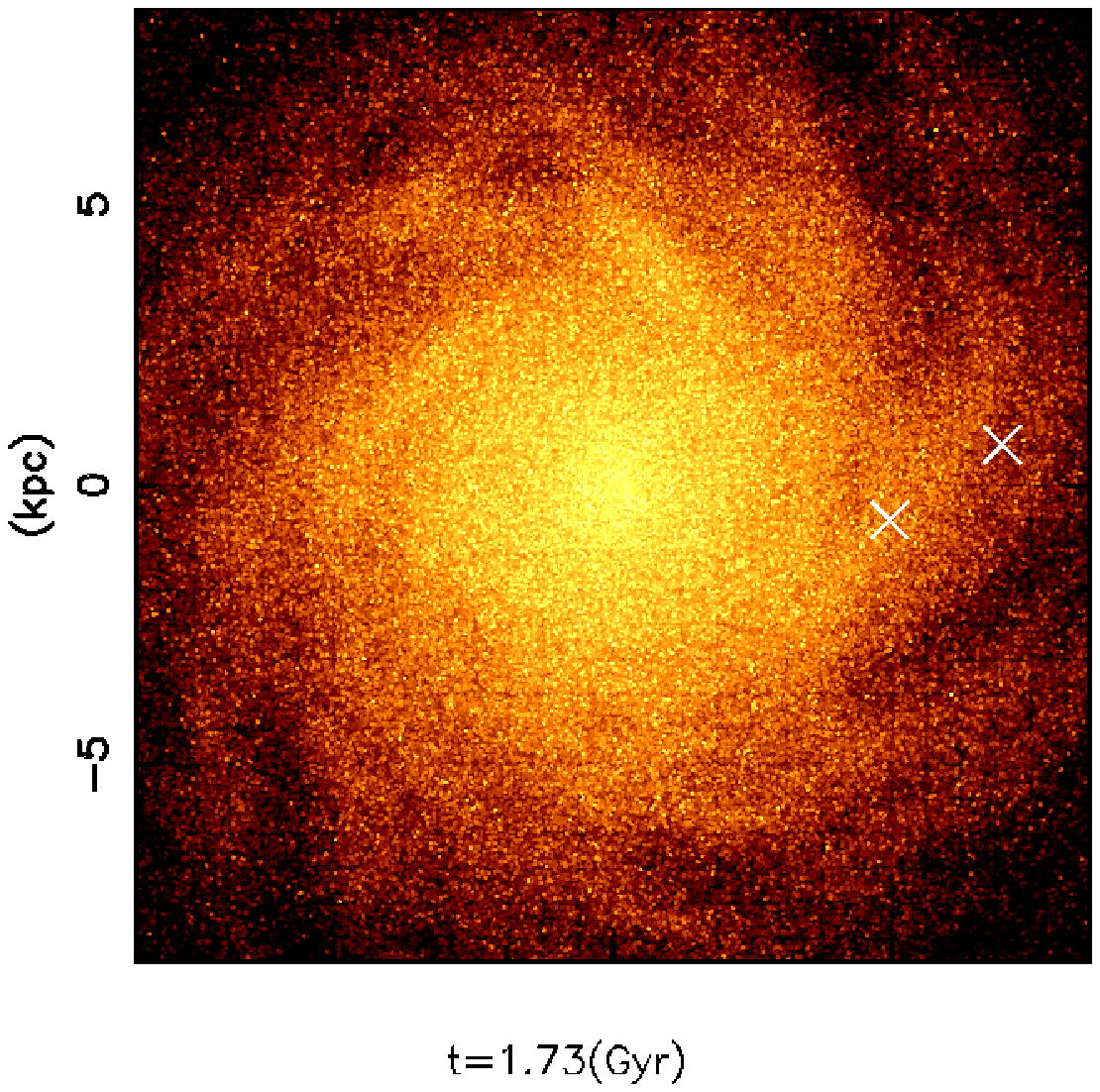}} \\ 
  
  \subfloat{\label{fig:t=1.79Gyr}\includegraphics[scale=0.44]{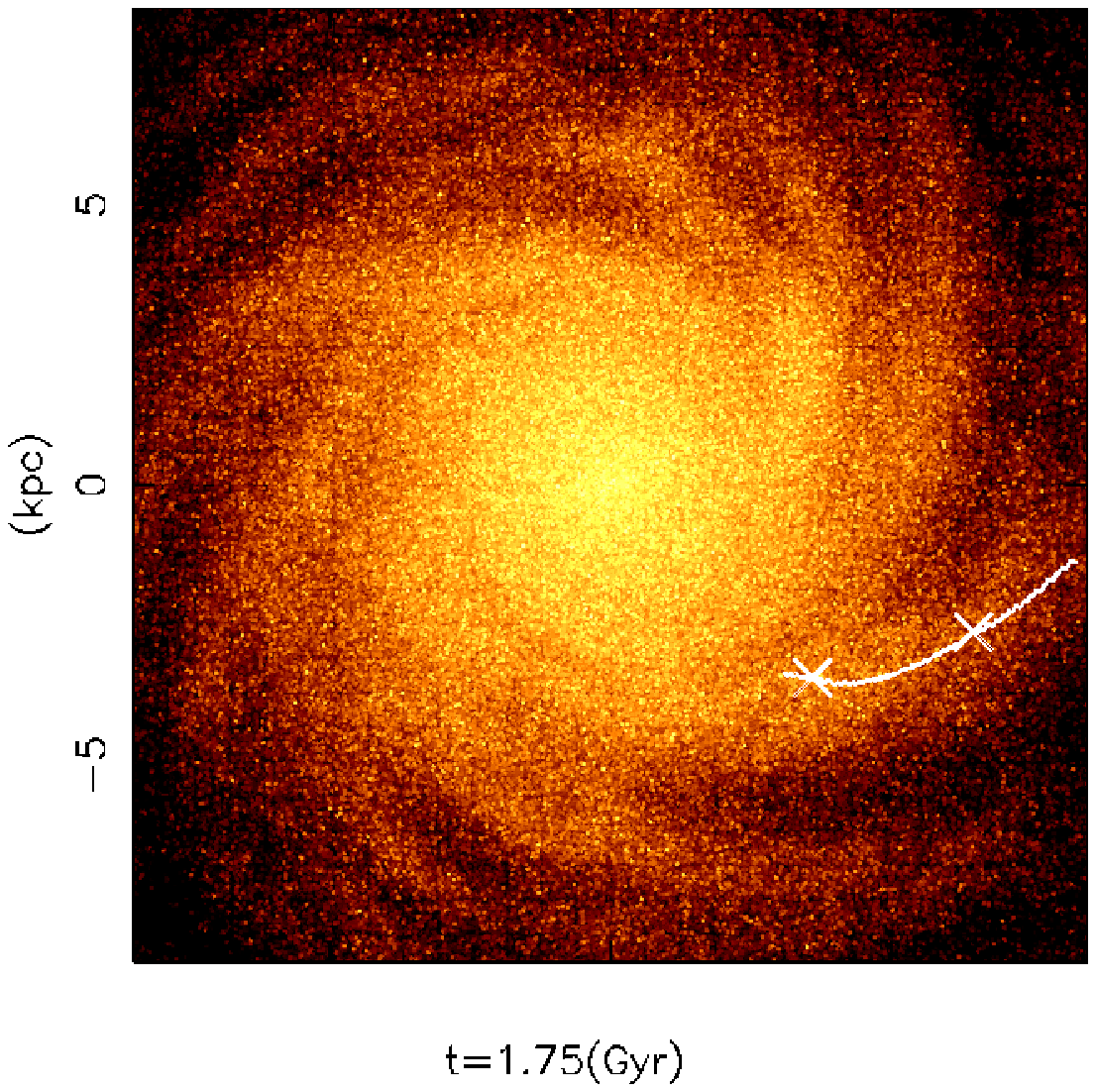}} \hspace{5.0mm} 
  \subfloat{\label{fig:t=1.75Gyr}\includegraphics[scale=0.44]{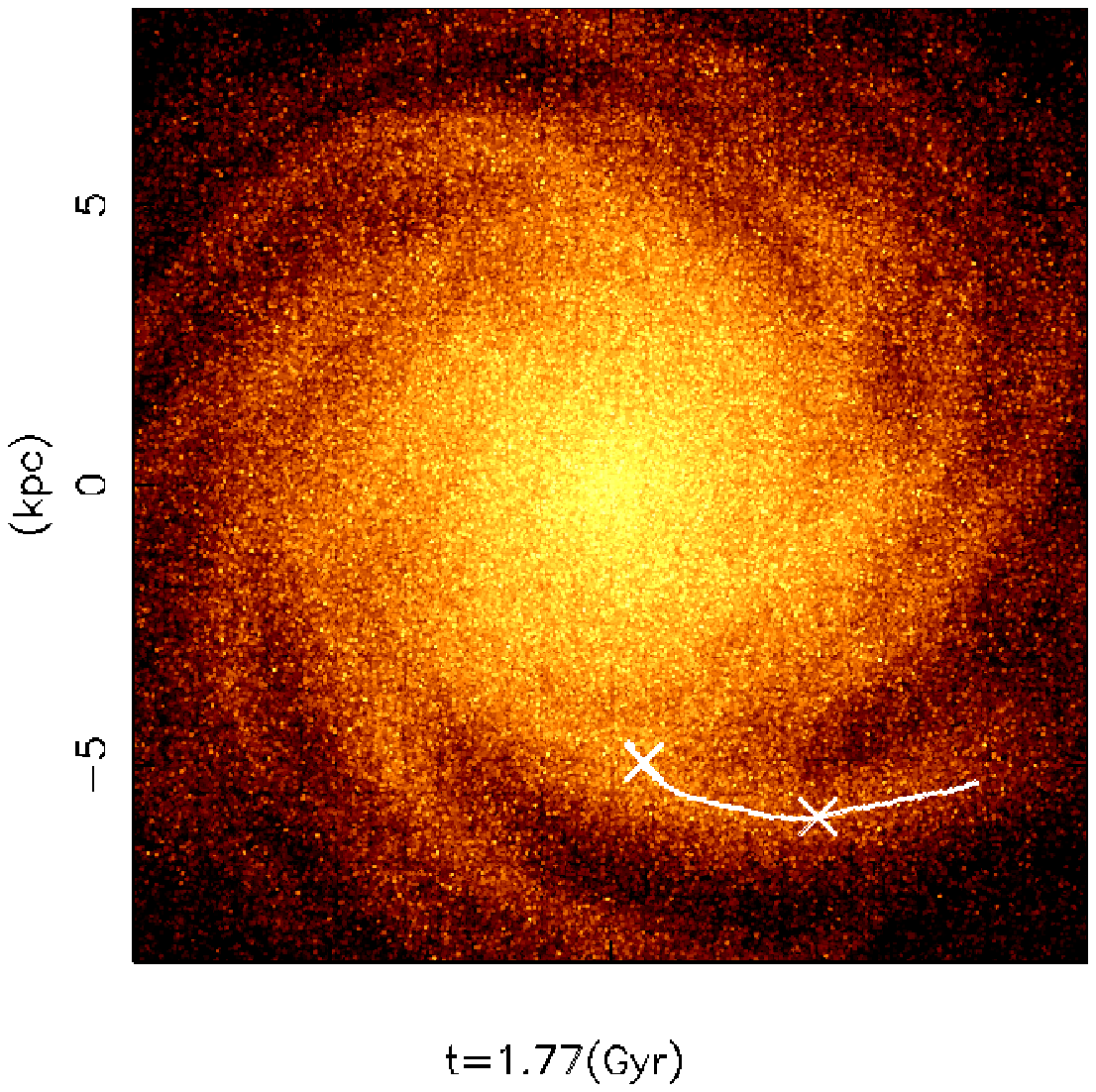}} \hspace{5.0mm} 
  \subfloat{\label{fig:t=1.77Gyr}\includegraphics[scale=0.44]{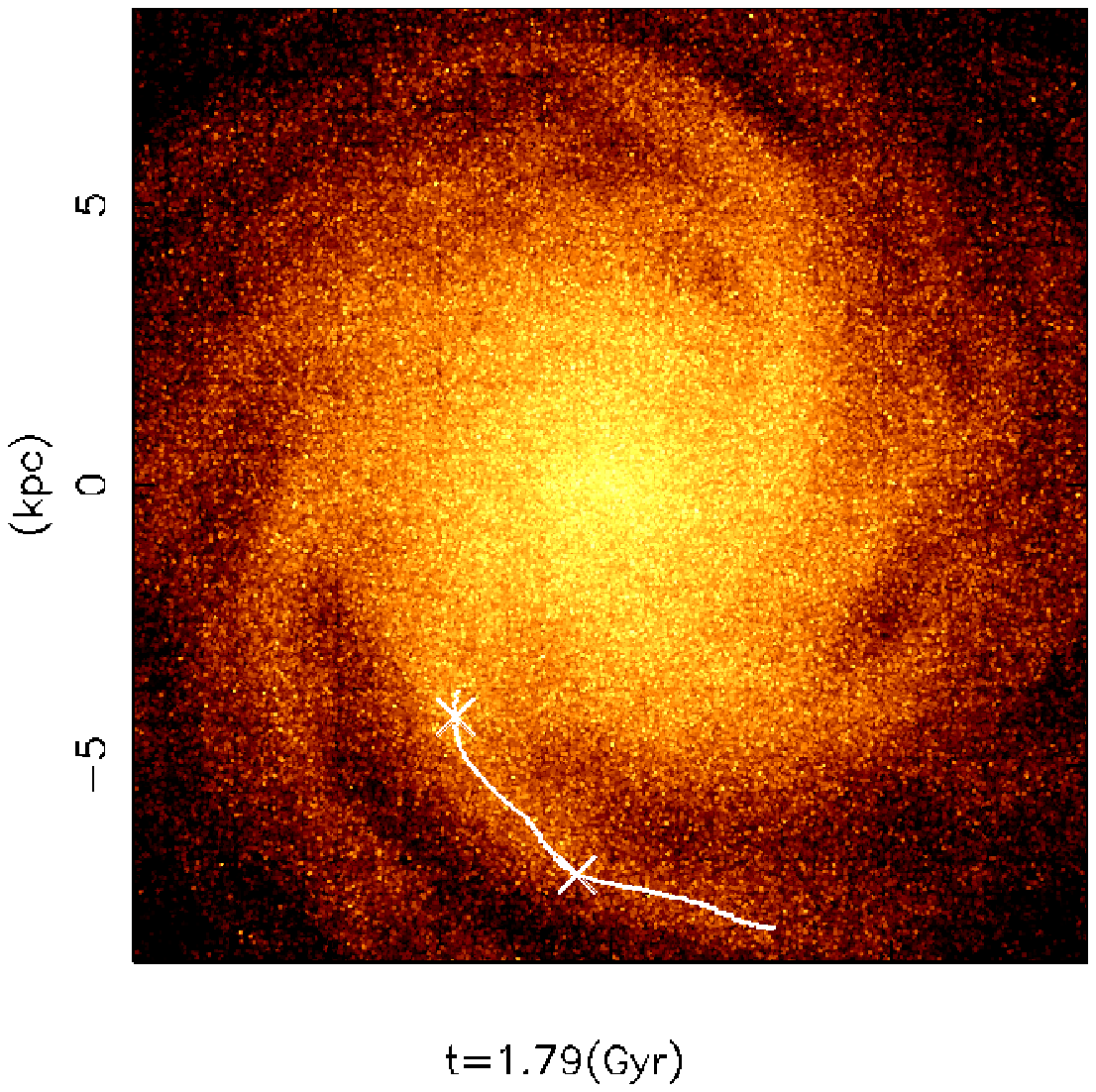}} \\ 
  
  \subfloat{\label{fig:t=1.79Gyr}\includegraphics[scale=0.44]{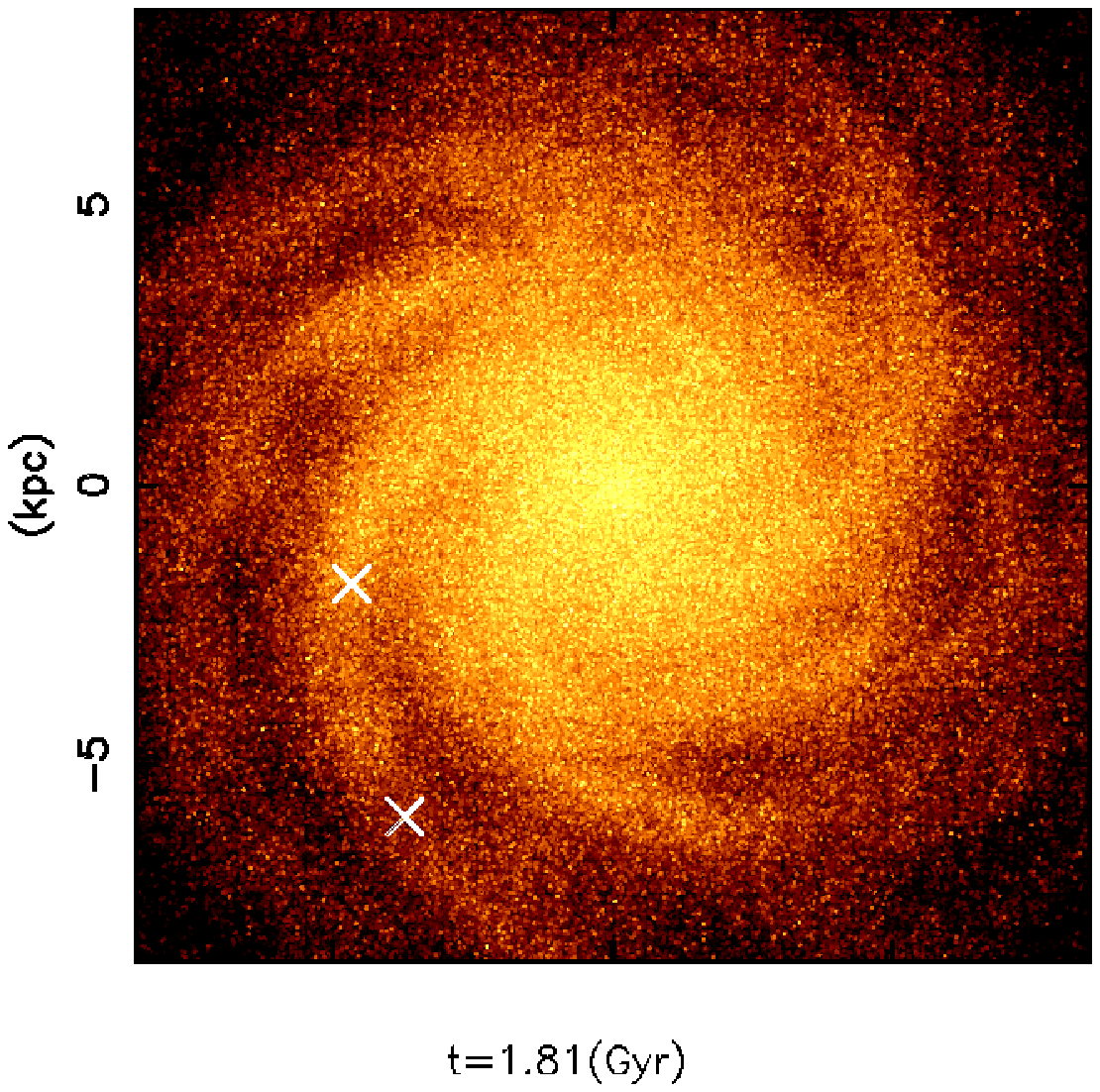}} \hspace{5.0mm} 
  \subfloat{\label{fig:t=1.75Gyr}\includegraphics[scale=0.44]{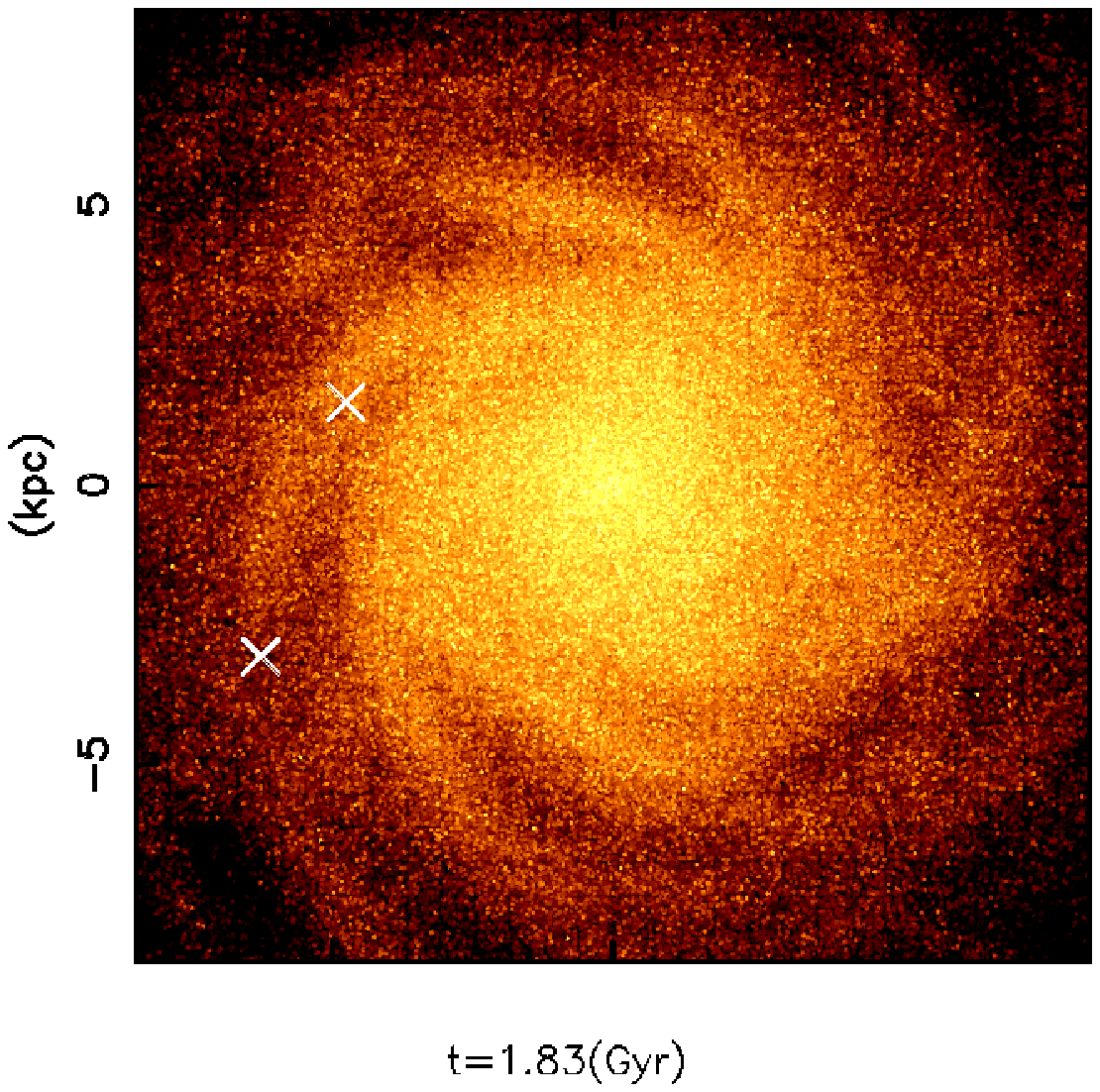}} \hspace{5.0mm} 
  \subfloat{\label{fig:t=1.77Gyr}\includegraphics[scale=0.44]{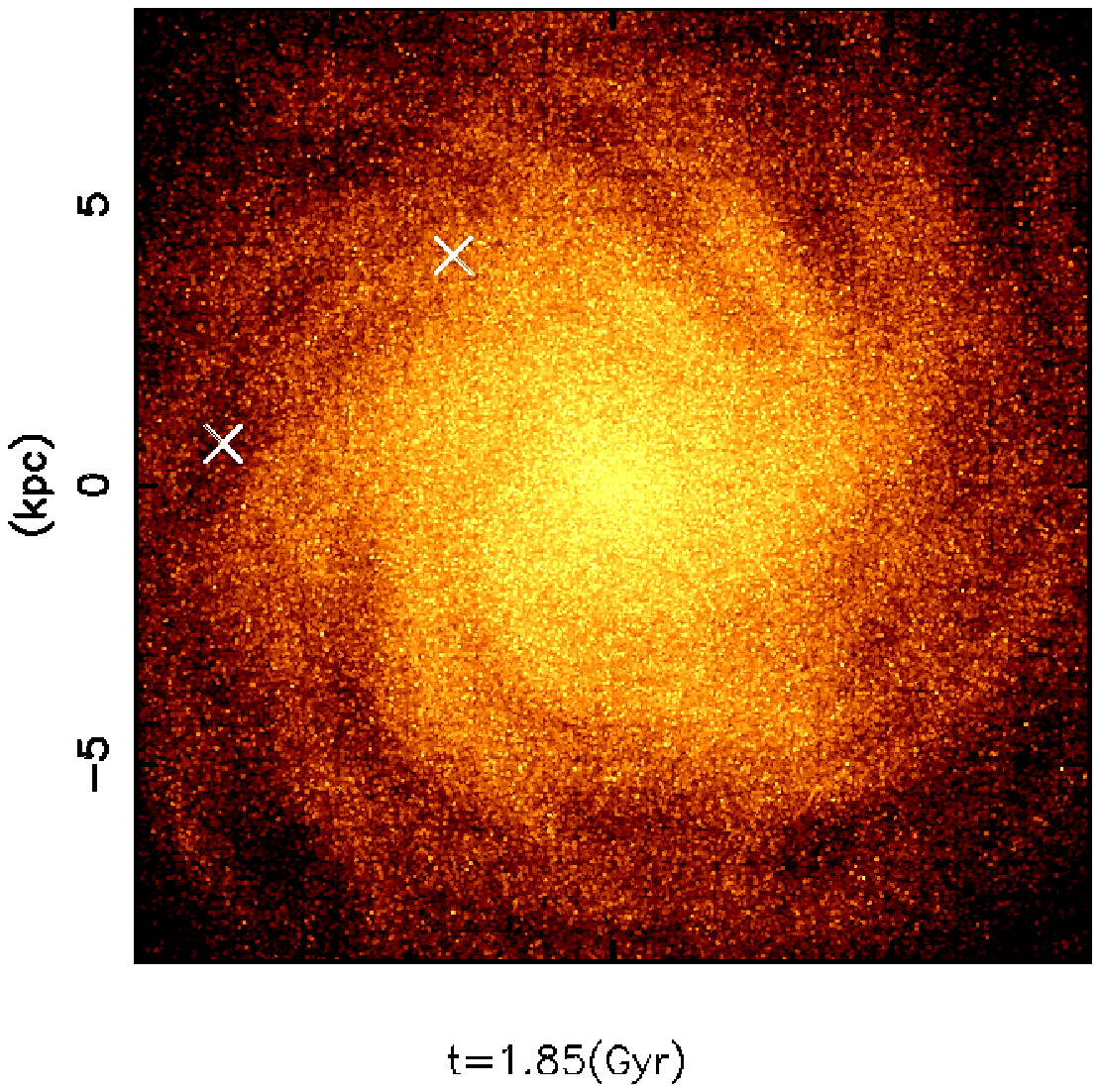}} \\ 
  
  \caption{Snapshots of the disc from face on view. The white line marks the position of the peak density of the fully formed spiral arm at the time indicated (see text for how this is determined). In the middle row, the white crosses at 5 kpc and 7 kpc radius indicate position of the peak density line at those radii. In the top and bottom rows, the crosses have been rotated from their positions in the middle panel (t$=1.77$ Gyr) by an angle derived from the angular rotation speed in Fig. \ref{omgp} for those radii. The purpose of the crosses is to guide the eye to the spiral arm from formation around t$=1.73$ Gyr, to its apparent destruction around t$=1.85$ Gyr. This indicates that the lifetime of the spiral arm is very short ($\sim 120$ Myr).}
 \label{snp1338}
\end{figure*}

\begin{figure}
\centering
\includegraphics[scale=0.42]{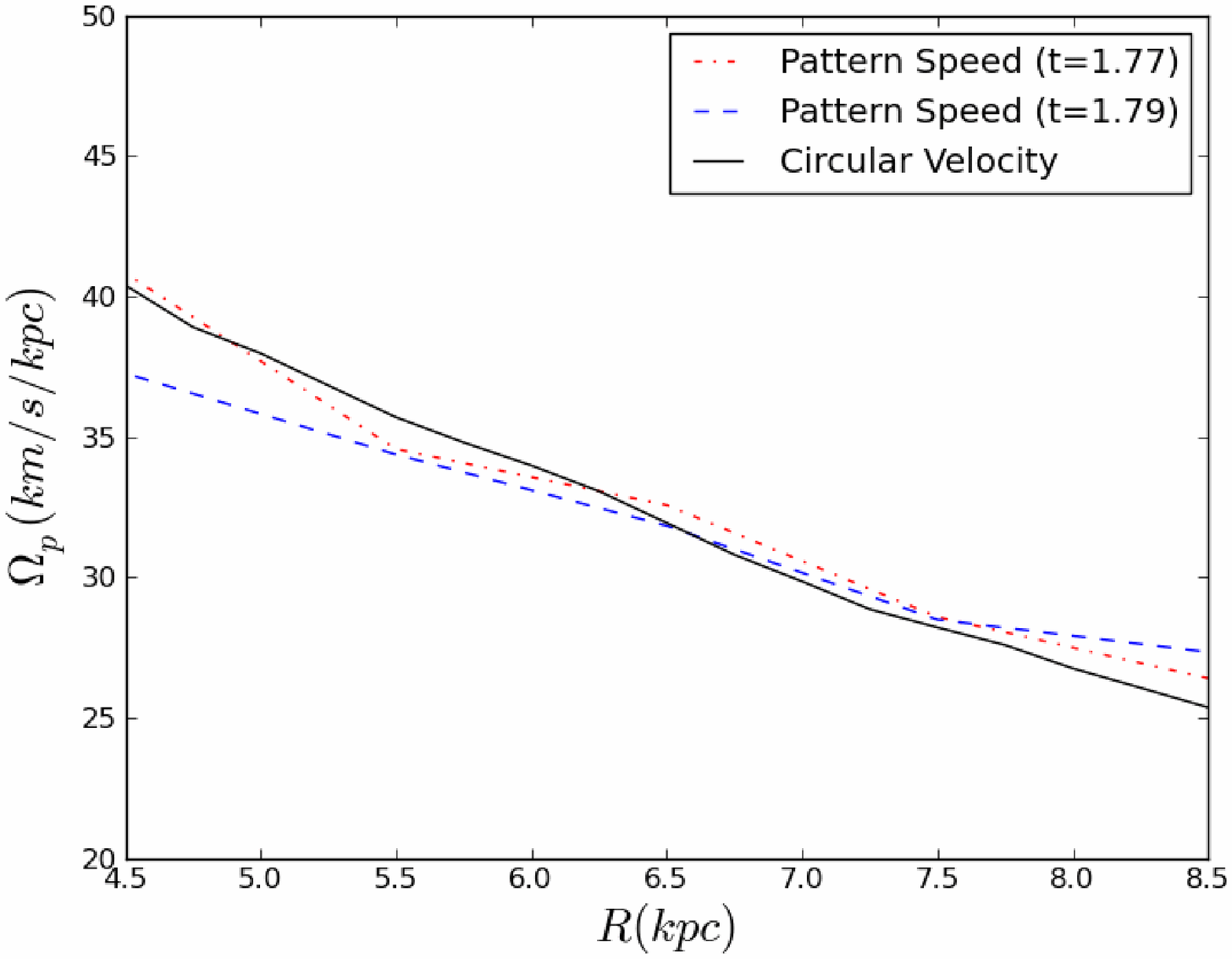} 
\caption{The pattern speed calculated between t$= 1.75$ and $1.77$ Gyr (dot-dashed red) and between t$=1.77$ and $1.79$ Gyr (blue dashed). The circular velocity at t= 1.77 Gyr shown in Fig. \ref{vcrotc} is also plotted (black solid). The pattern speeds agree well with each other, and exhibit a decreasing trend that traces the circular velocity very closely over the radial range shown. Similar trends are found at other time steps with different arms.}
\label{omgp}
\end{figure}

\subsection{Pattern Speed and Radial Migration}

\begin{figure*}
\begin{center}

  \subfloat{\includegraphics[scale=0.55]{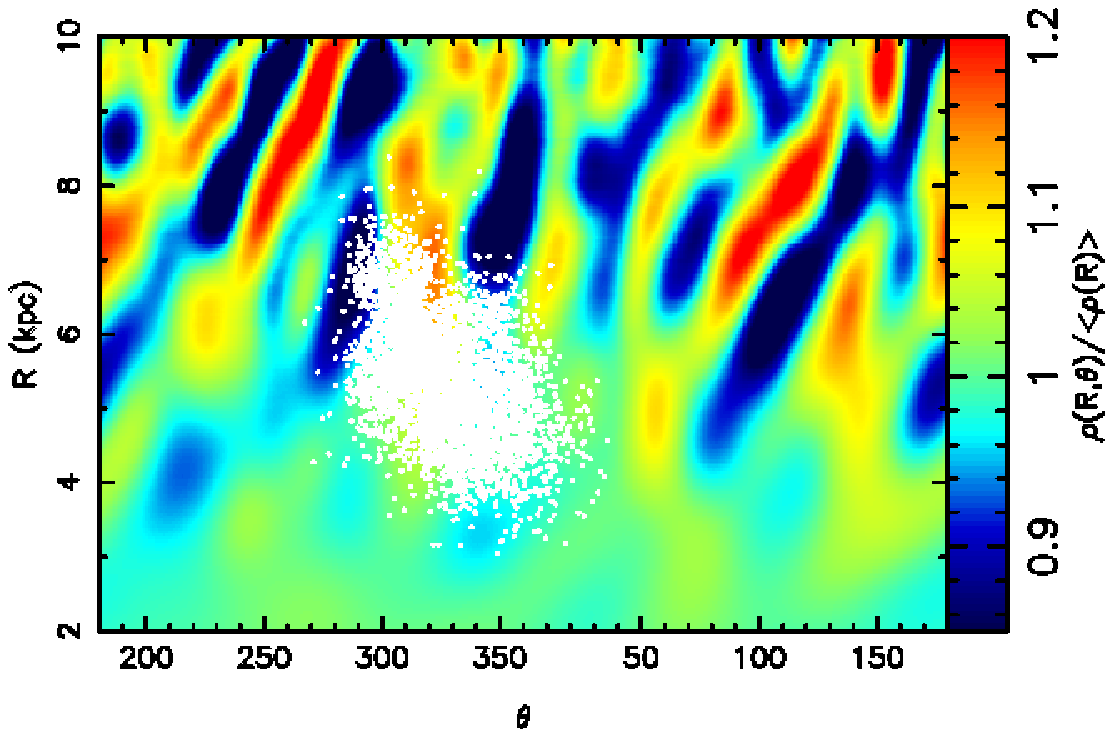}}  
   \subfloat{\includegraphics[scale=0.55]{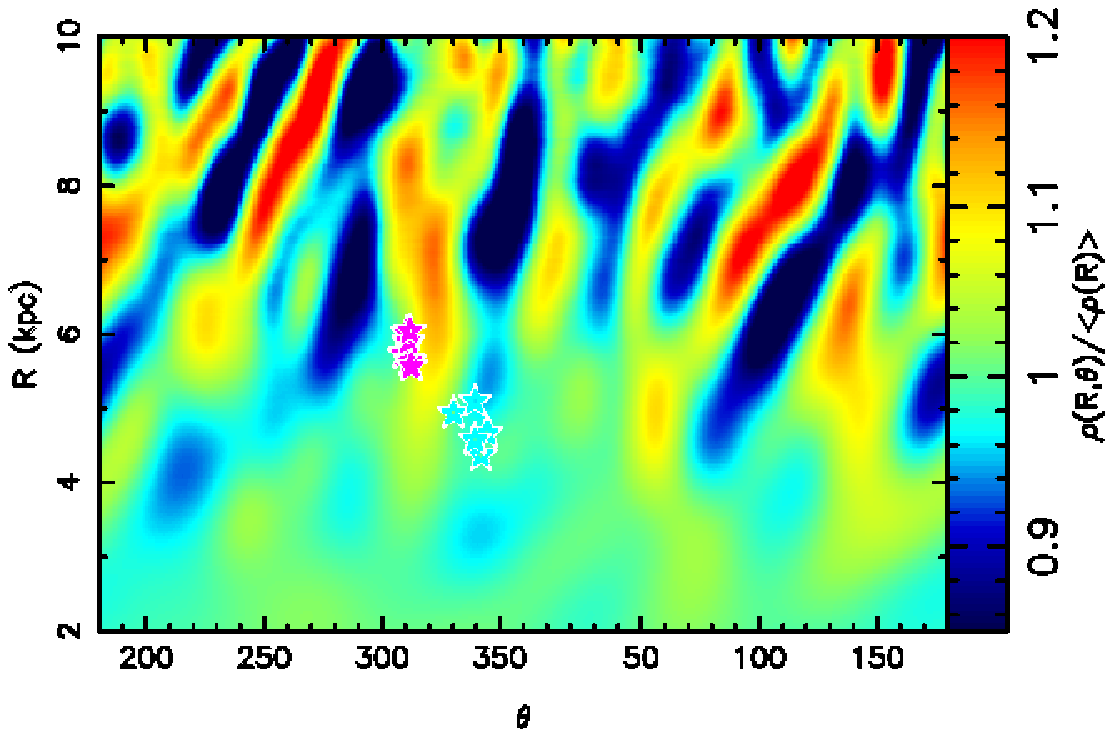}}  
    \subfloat{\includegraphics[scale=0.35]{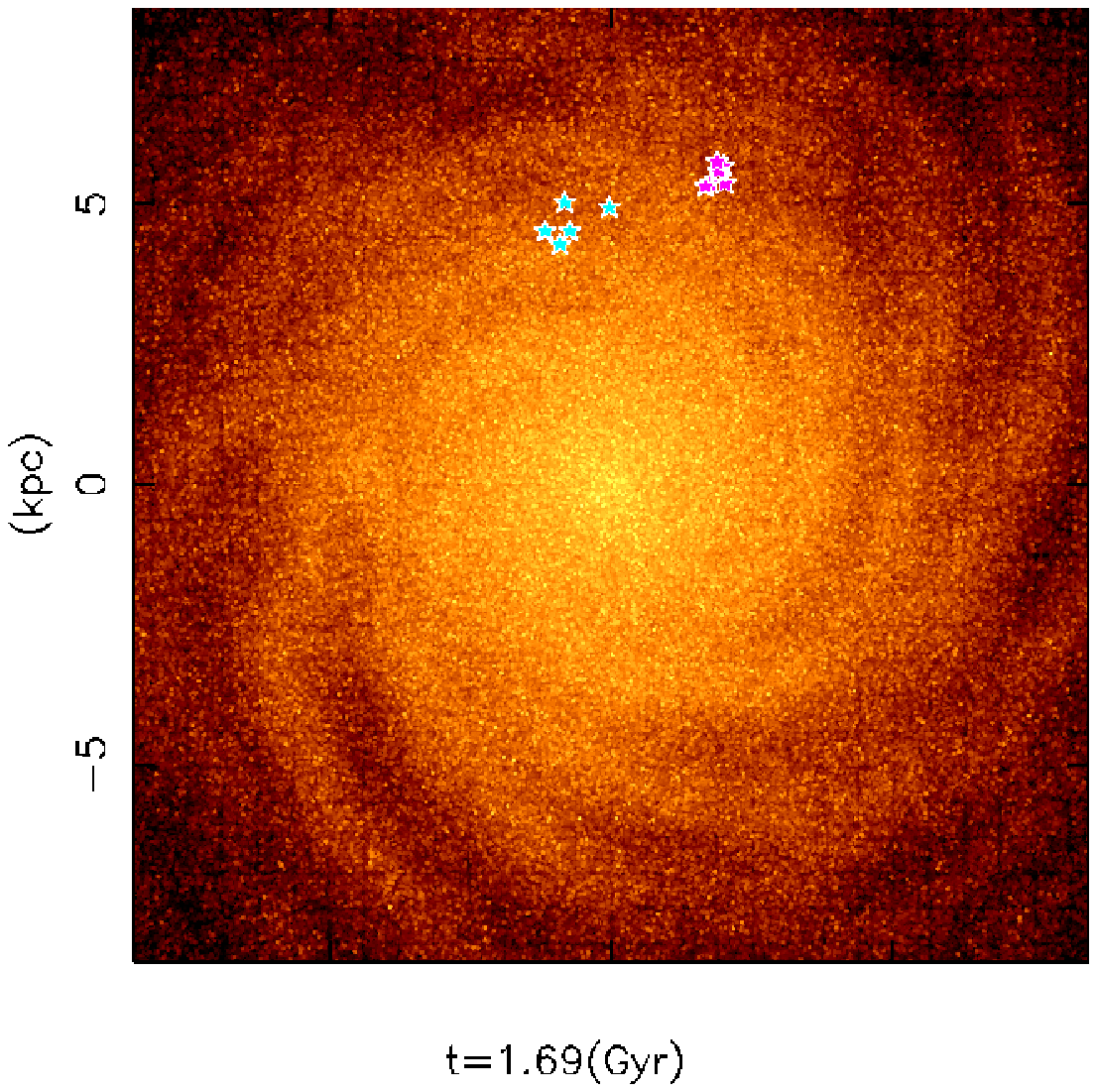}} \\  

  \subfloat{\includegraphics[scale=0.55] {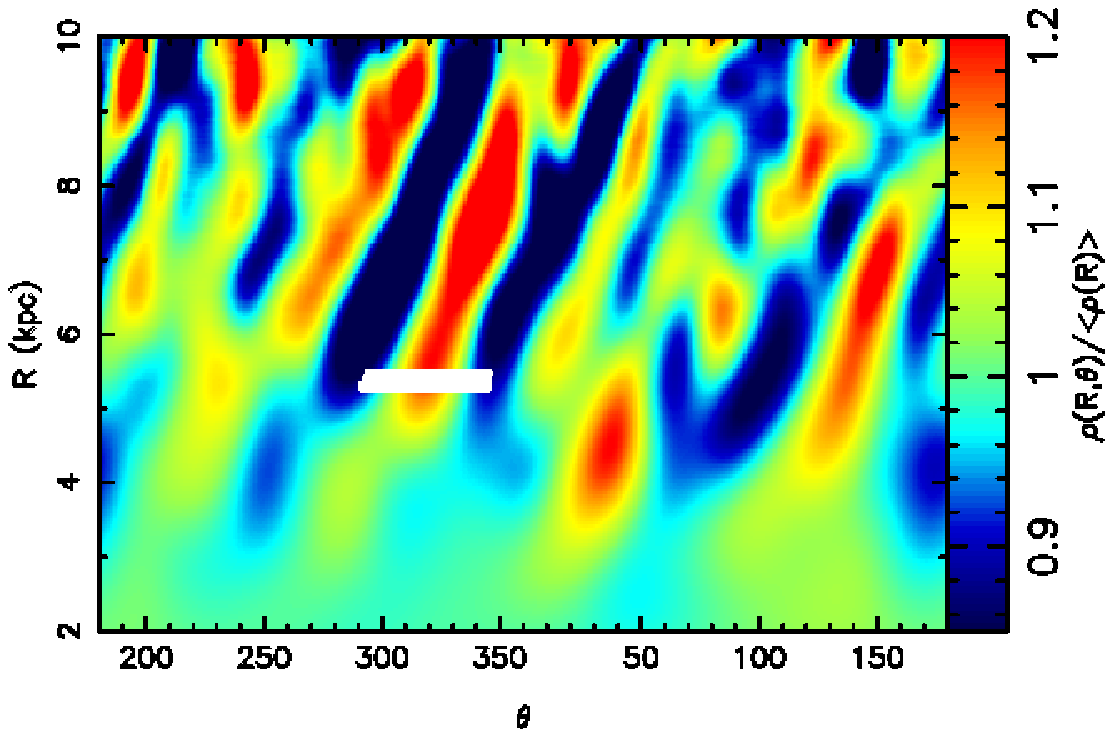}}  
   \subfloat{\includegraphics[scale=0.55] {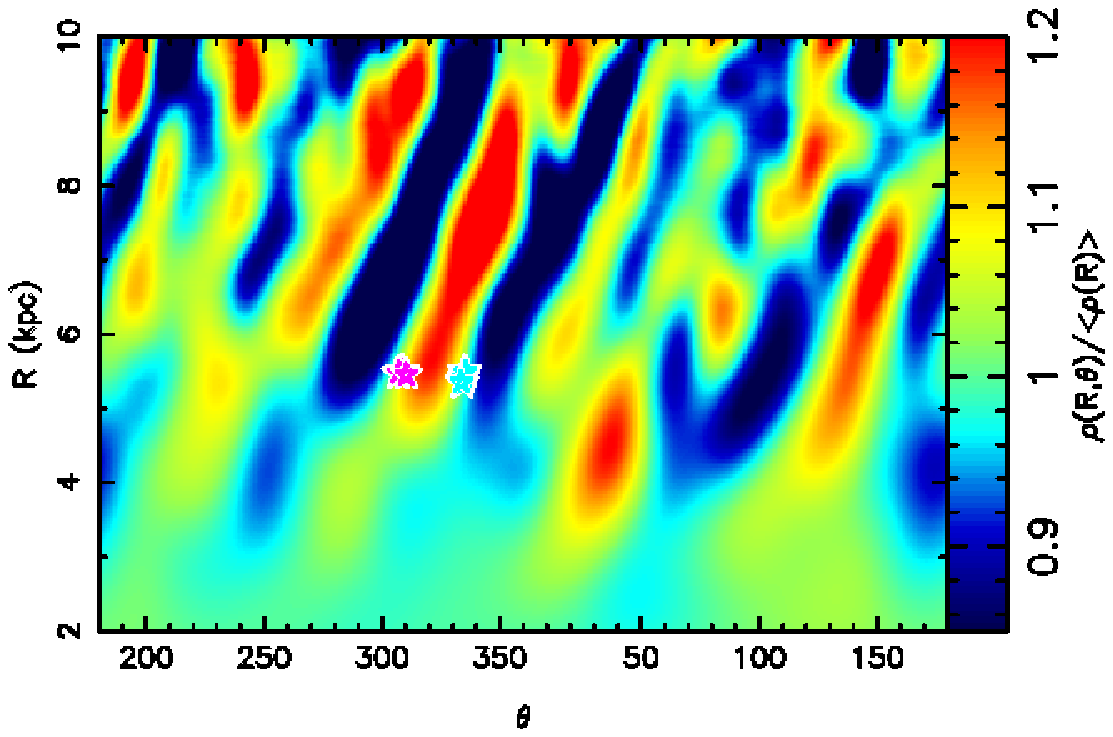}}  
    \subfloat{\includegraphics[scale=0.35]{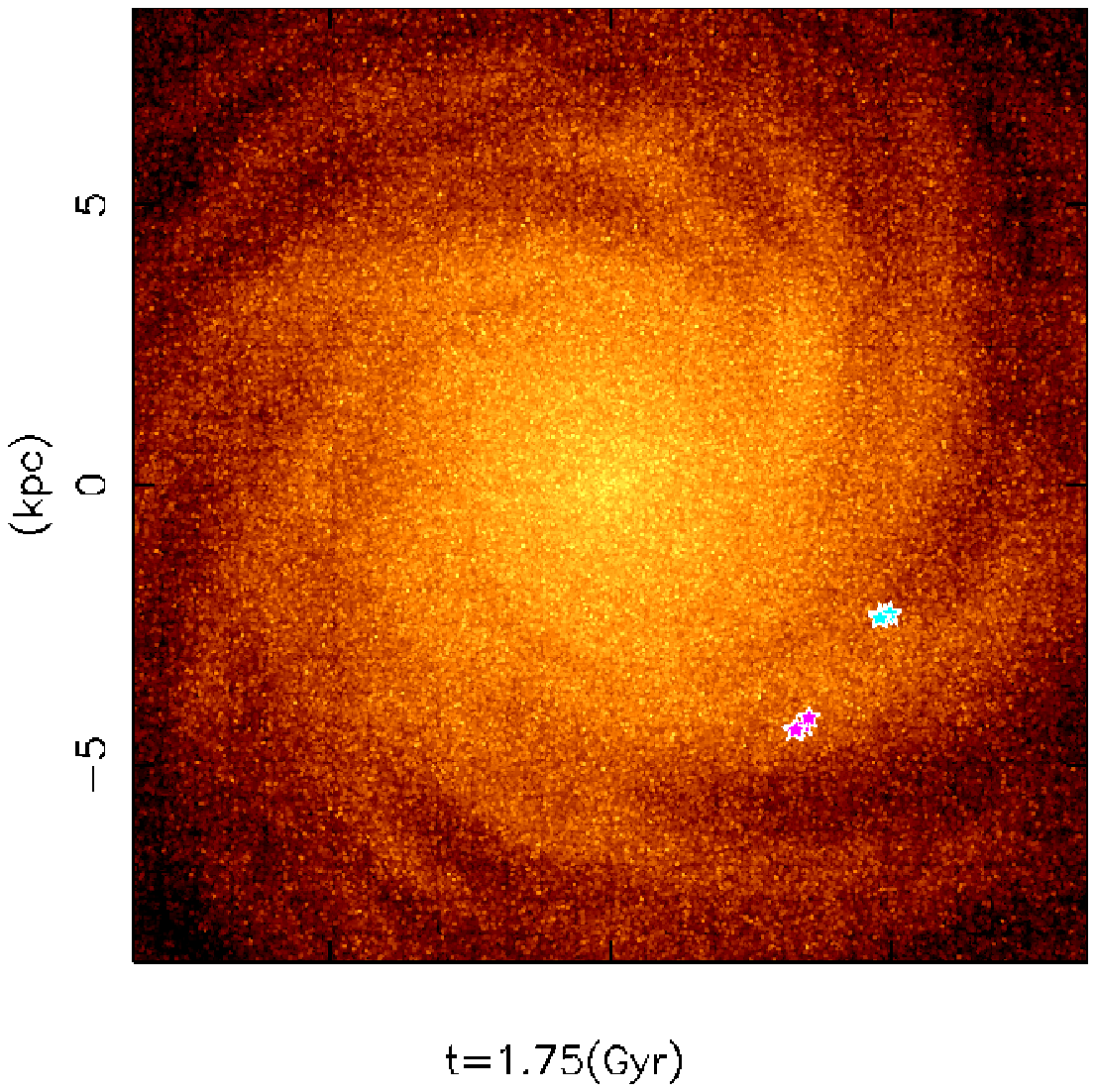}}\\     

\subfloat{\includegraphics[scale=0.55] {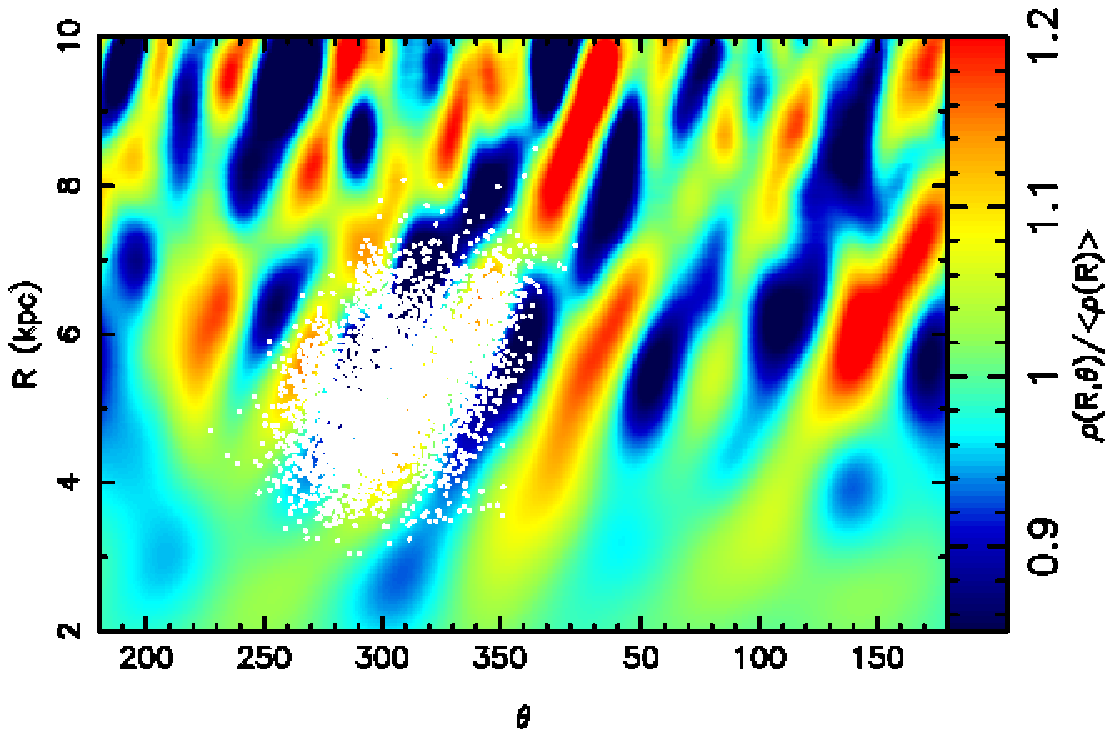}}  
\subfloat{\includegraphics[scale=0.55] {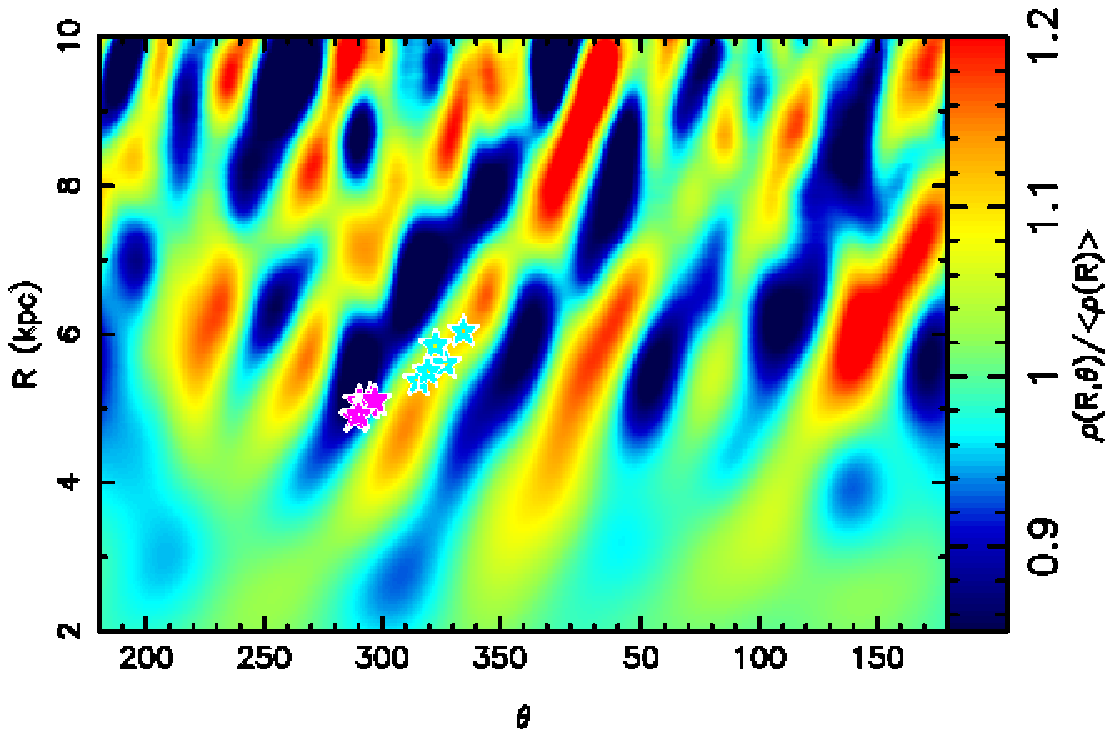}}   
\subfloat{\includegraphics[scale=0.35]{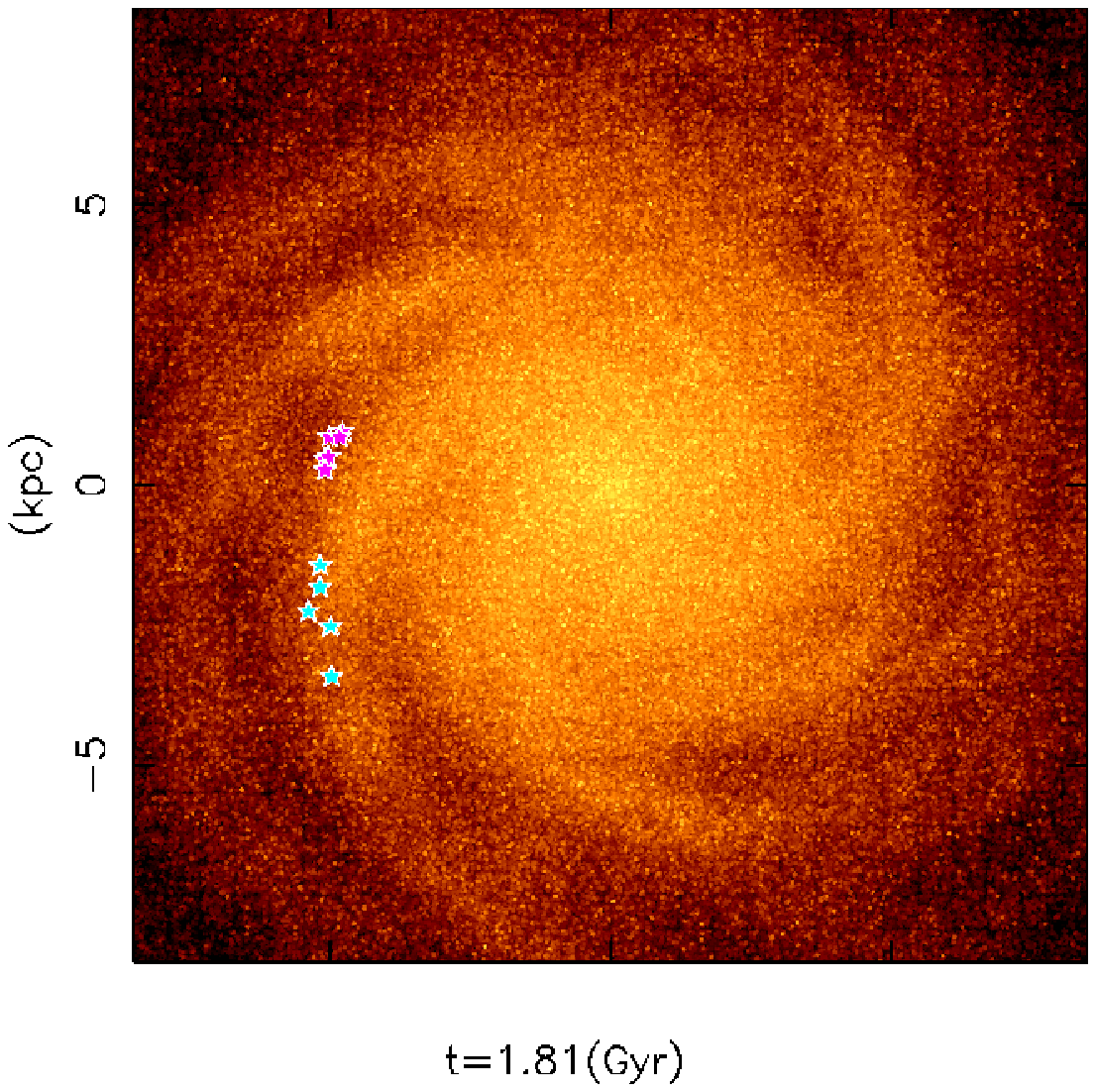}} \\     

\caption{Left and Middle columns: the smoothed normalised density distribution (colour map) in the azimuthal angle - radius plane. Right column: the corresponding snapshots of the disc (the time referring to each row is indicated under each panel). In the left column, the particles (white dots) selected at 5.5 kpc at t= 1.75 Gyr are highlighted. In the middle and right columns the extreme migrators (see text) in the sample are highlighted by cyan (particles that migrate toward the outer radii) and pink (particles that move toward the inner radii) stars. Note that the coordinate of the angle in the left and middle panels at the top and bottom rows are shifted by the amount corresponding with $\Omega = 40 \rm kms^{-1}kpc^{-1}$, to keep the highlighted particles around the central region of each panel.}
\label{5kpcplots}
\end{center}
\end{figure*}

The middle row of panels in Fig. \ref{snp1338} shows three consecutive snapshots of the model galaxy, where the white lines depict the peak density of the chosen spiral arm at each radius. This is calculated by first creating a smoothed normalised density distribution in the $R$-$\theta$ plane, as in the left and middle columns of Fig. \ref{5kpcplots}. At each radial bin, we locate the azimuth of maximum density. In this way, we traced the arm in the radial range between 4.5 and 8.5 kpc, and so we focus our analysis on this radial range. The pattern speed, $\Omega _p$, is then easily found from the azimuthal offset of the peak densities between the time steps. 

The angular pattern speed measured from the snapshots in Fig. \ref{snp1338} (middle row) is plotted in Fig. \ref{omgp}. It is seen to decrease with radius, such that it almost equals the circular velocity of stars (Fig. \ref{vcrotc} also shows circular velocity in angular terms at $t=1.77$ Gyr for reference) in the disc at all radii. This indicates that this spiral arm feature is co-rotating, and is therefore unlikely to be long-lived. This is confirmed from the snapshots of our simulation shown in the top and bottom rows of Fig. \ref{snp1338}, which show that the arm starts developing around t=$1.73$ Gyr and is winding and disrupted around t=$1.85$ Gyr. Hence the lifetime of the arm is about $120$ Myr and the arm is indeed transient. Bifurcations and breaks in the spiral arm features are seen regularly, which occur as the arms wind up. 
\citet{WBS11} also find that the spiral arm features in their simulations are co-rotating, winding and short-lived.


\begin{figure*}
 \centering
  \subfloat{\label{left}\includegraphics[scale=0.43]{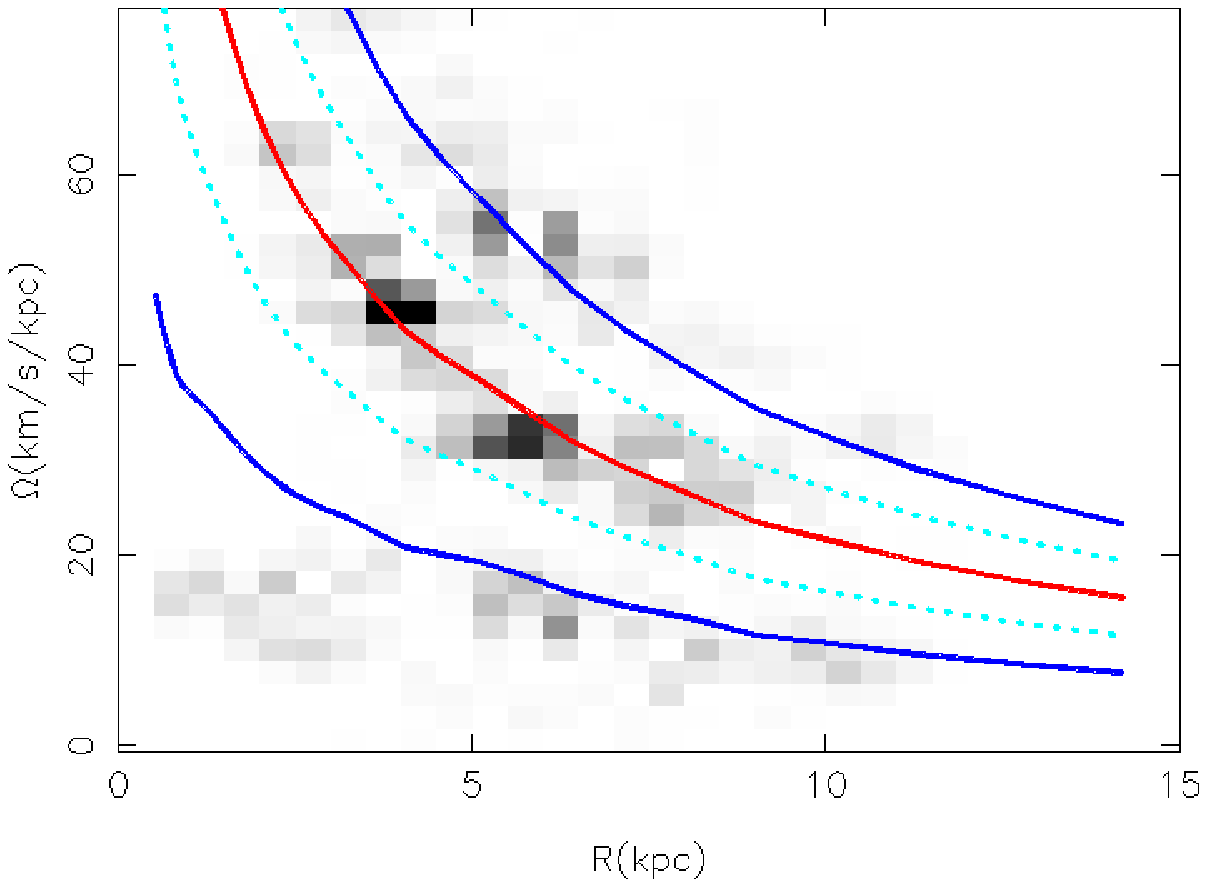}} \hspace{-8.0mm} 
  \subfloat{\label{middle}\includegraphics[scale=0.43]{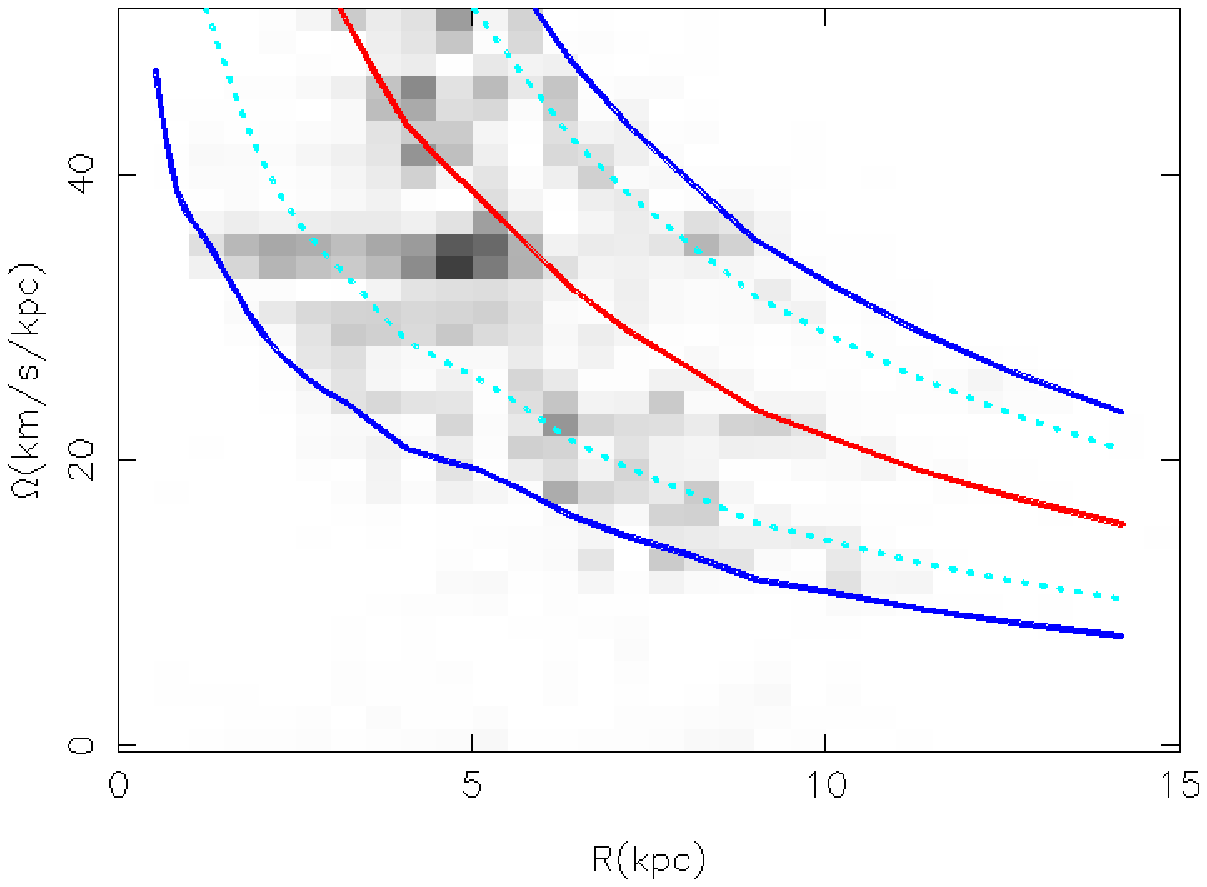}} \hspace{-8.0mm} 
  \subfloat{\label{right}\includegraphics[scale=0.43]{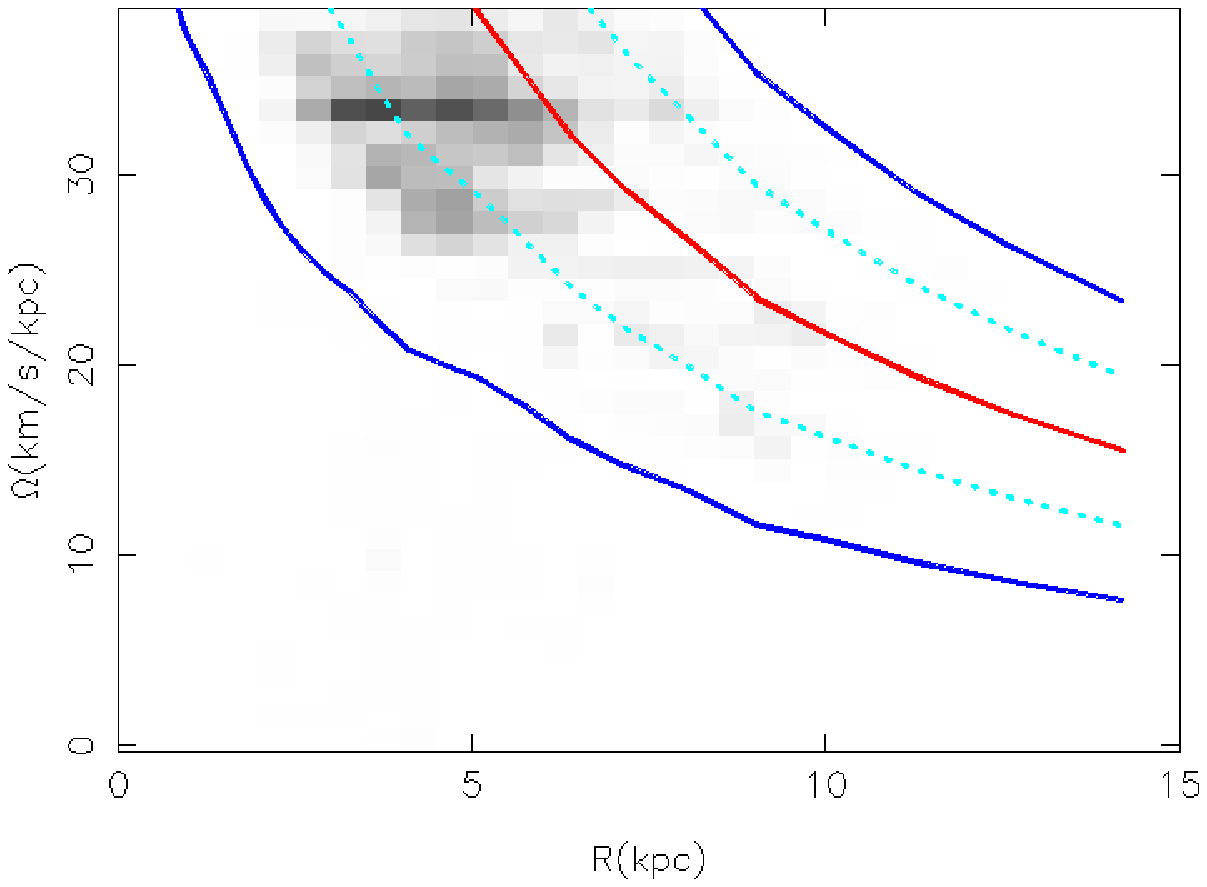}} 
  \caption  {Spectrograms of the $m=2, 3$ and $4$ Fourier components in a time window spanning 1.28 Gyr centred on $t \sim 1.77$ Gyr. The plots show the power in frequencies ranging from 0 to the Nyquist frequency as a function of radius. \emph{Left}: The spectrogram of the $m=2$ component. Also marked is the circular velocity line (solid red), the inner and outer Lindblad resonances (solid blue) and the 4:1 resonances (dotted cyan). \emph{Middle}: The same as left panel but for $m=3$ Fourier component, and 3:1 resonances shown (dashed cyan). \emph{Right}: The same as left panel but for $m=4$ Fourier component.}
 \label{spectrog}
\end{figure*}

\begin{figure*}
 \centering
  \subfloat{\label{fpk33}\includegraphics[scale=0.417]{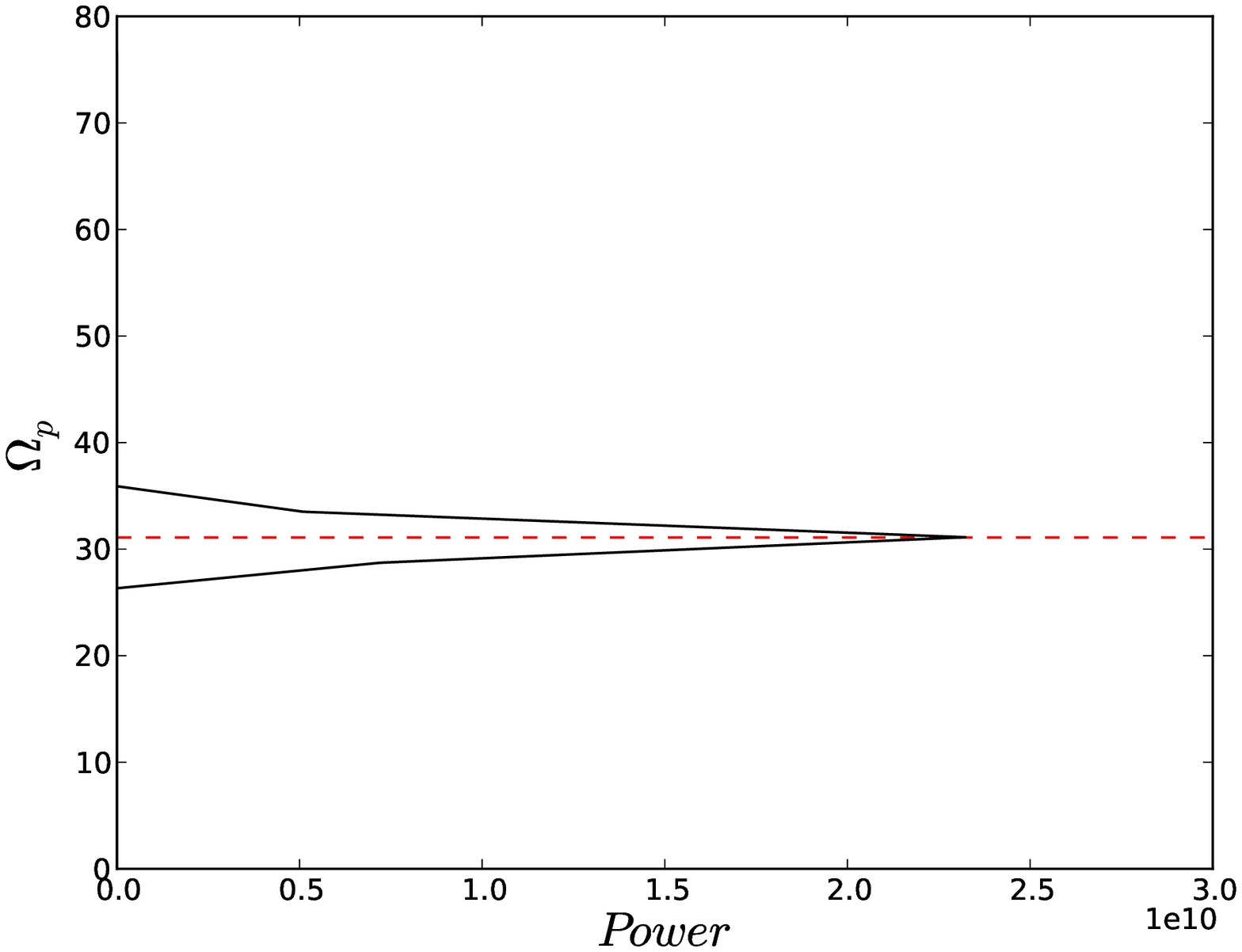}} \hspace{0.0mm} 
  \subfloat{\label{fpk25}\includegraphics[scale=0.417]{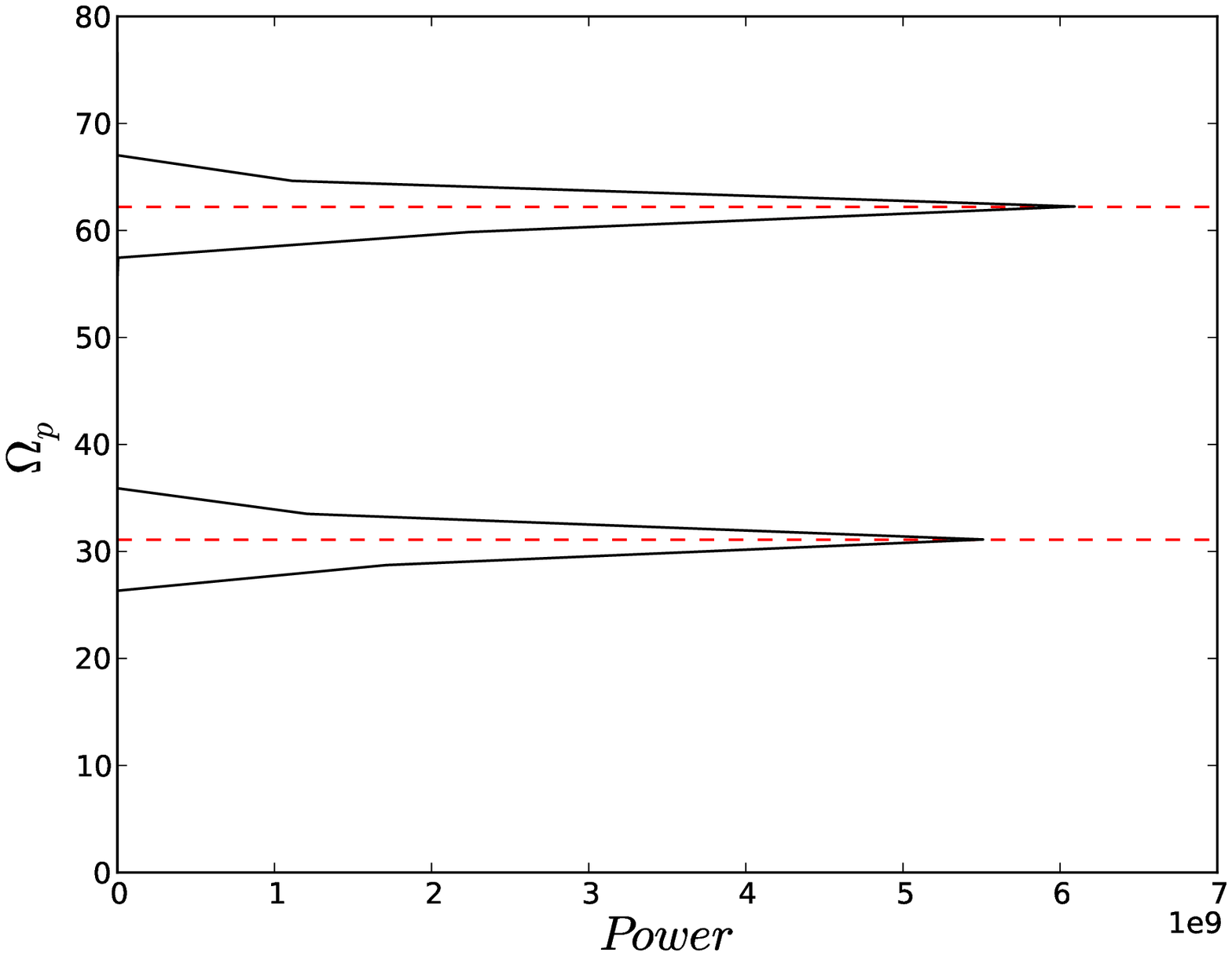}} \hspace{0.0mm} \\ 
  \subfloat{\label{fpk33v}\includegraphics[scale=0.417]{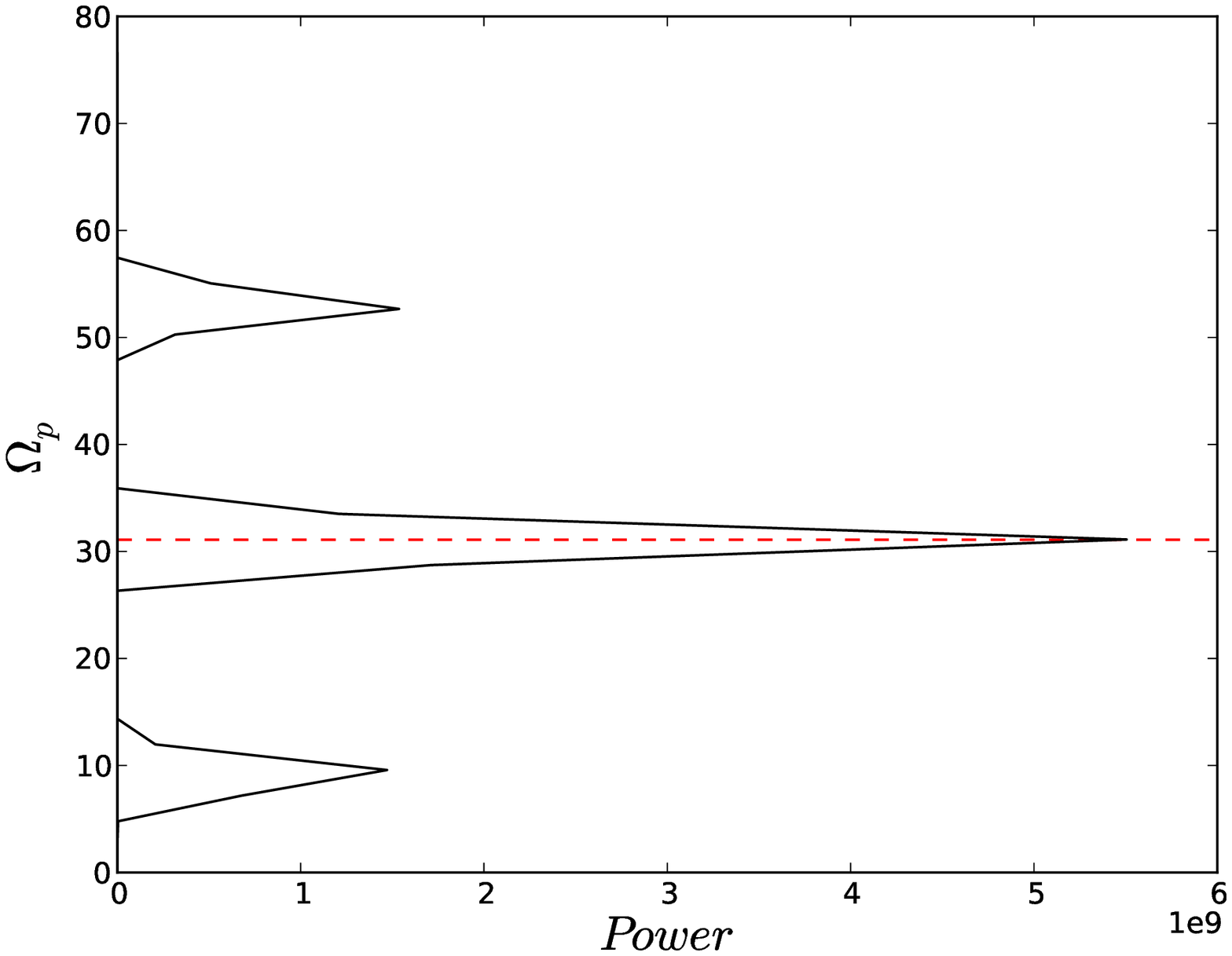}} \hspace{0.0mm} 
  \subfloat{\label{fpk25v}\includegraphics[scale=0.417]{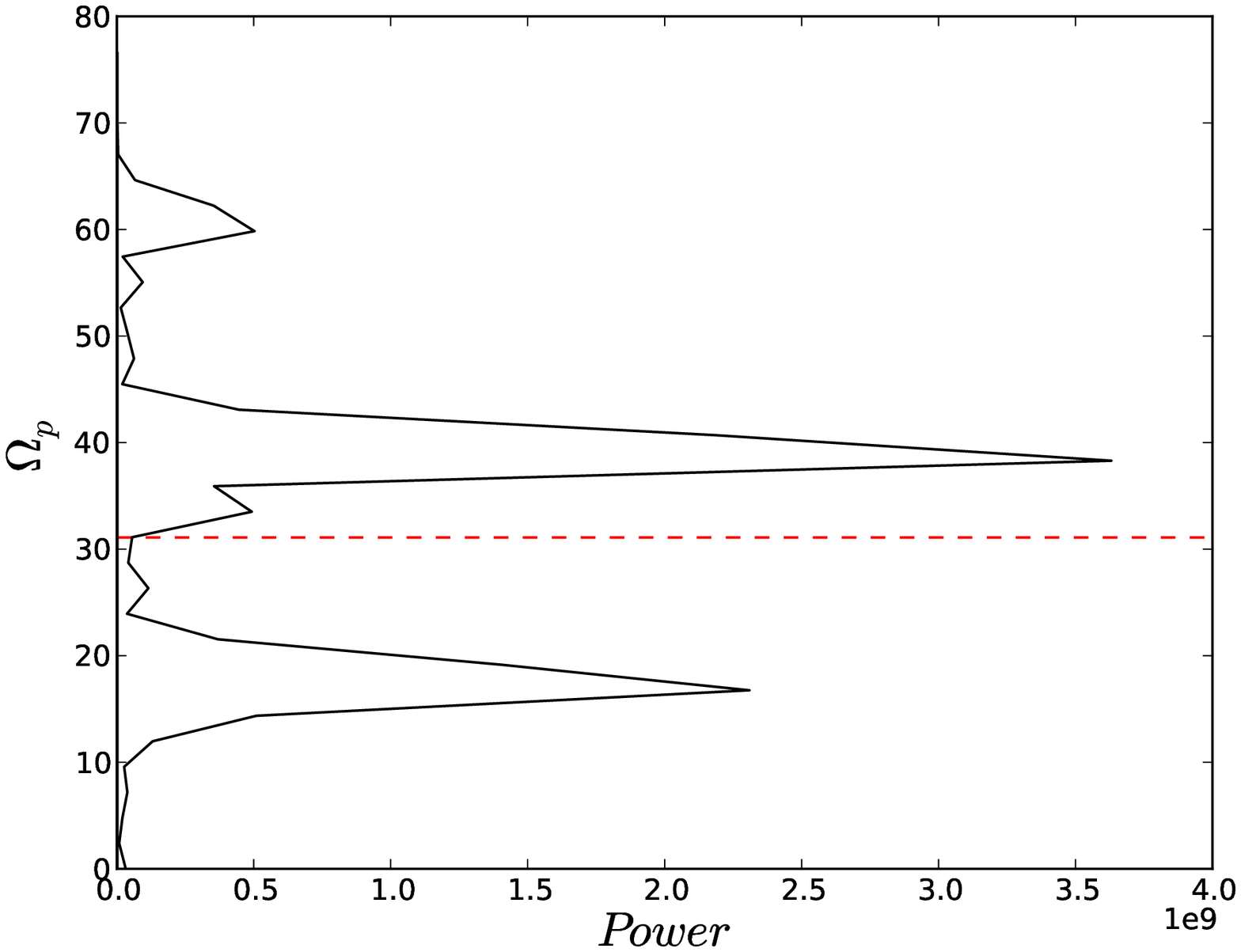}} \hspace{0.0mm} \\ 
  \caption  {Input pattern speeds of the $m = 2$ wave modes (dashed red line) and the power recovered from the spectrogram analysis (solid black line). \emph{Top left}: Single $\Omega _p = 31$ km s$^{-1}$ kpc$^{-1}$ is input and correctly extracted. \emph{Top right}: Two wave modes of $\Omega _p = 31$ and $62$ km s$^{-1}$ kpc$^{-1}$ are input. \emph{Bottom left}: Same as top left, but the amplitude of the wave mode is varied. \emph{Bottom right}: Same as bottom left, but random offsets of phase angle are implemented. See text for details.}
 \label{fconv}
\end{figure*}

As an additional test, and to assist intercomparison with previous studies, we compute spectrograms for $m=2, 3$ and $4$ modes, following \citet{QDBM10}. At each time step, which spans a time period of 1.28 Gyr centred on $t \sim1.77$ Gyr, and each radial bin, we calculate 

\begin{equation}
W_c(r,t,m) = \sum_i \cos{(m \theta_i)},
\end{equation}

\begin{equation}
W_s(r,t,m) = \sum_i \sin{(m \theta_i)},
\end{equation}
where $\theta _i$ is the angle at the position of the particles within the radial bin. We calculate the amplitude of each mode as a function of radius and find that all modes show similar strength at $t = 1.77$ Gyr, hence there is no one mode that dominates over the others. We then compute the Fourier transform in the above period,
\begin{equation}
\tilde{W}(\omega ,r,m)= \int_{T_1}^{T_2} [W_c(r,t,m)+i W_s(r,t,m)] e^{i \omega t} h(t) dt  
\end{equation}
where $h(t)$ is the Hanning function. We compute the power at each frequency ranging from zero to the Nyquist frequency, which is shown in Fig. \ref{spectrog}. The spectrogram analysis shows us the significance of the pattern speeds of the wave modes (which we call \emph{mode pattern speed}), which can be different from what is shown in Fig. \ref{omgp} (i.e. the pattern speed of the spiral arm feature), if multiple wave modes interfere with each other. All three modes, especially the $m = 2$ mode, show that there is some power in the mode pattern speed that overlaps the circular velocity at many radii. There are also several horizontal features which could mean that there are a number of wave modes with constant, but different pattern speeds (\citet{Se11} and references therein), that span different radii, and may be constructively and destructively interfering with each other. For example, the $m= $ 2 mode could be interpreted as having two pattern speeds ($\sim 30$ and $45$ km s$^{-1}$ kpc$^{-1}$), with a faster inner pattern and a slower outer pattern, as suggested by \citet{QDBM10}; \citet{SL89}; \citet{MT97}. Fig. \ref{spectrog} also indicates the 1:2, 1:3 and 1:4 Lindblad resonances, i.e. $\Omega_p=\Omega \pm \kappa/m$, for $m=2, 3$ and $4$ respectively. 

We note however some caution with regard to the spectrogram analysis method when applied to transient, variable amplitude wave modes. We construct a toy model in which we set a base density of stars and distribute them randomly in a ring. We then add a small fraction ($2.5 \%$) of those in the ring at $\theta = 0, \pi$ to mimic the $m=2$ modes. We impose a single, constant rotational frequency on the particles, and perform the same spectrogram analysis as above. Fig. \ref{fconv} (top left panel) shows the case assuming $31$ km s$^{-1}$ kpc$^{-1}$ rotation, for which this frequency is correctly extracted. The top right panel shows the case for two overlapping wave modes of $31$ and $62$ km s$^{-1}$ kpc$^{-1}$ rotation speeds, and both are shown to be extracted. 

However, the amplitude of the wave mode can evolve and/or disappear as demonstraced in \citet{Se11}. To demonstrate the effect of variable amplitude, we return to a single pattern speed of $31$ km s$^{-1}$ kpc$^{-1}$, and repeatedly increase and decrease the number of $m = 2$ mode particles using a sinusoidal curve from $0$ to $\pi$, that spans over $120$ Myr, which is roughly consistent with the lifetime of the arm feature in Fig. \ref{omgp}. The effect is shown (Fig. \ref{fconv} bottom left panel) to create three peaks, with one at the input frequency, and two others: one either side of the real one.

The appearance of new mode can shift the azimuthal position, which we model as randomly shifting the azimuthal position of the mode particles within $0$ to $\pi$, every $120$ Myr. This effect plus the variable amplitude is shown in the bottom right panel. This produces several peaks, none of which corresponds to the real input frequency. Although this is a simple exercise, and our aim is not to explore many possible cases of transient wave modes, this demonstrates that especially for transient waves modes, there is a possible danger that the spectrogram may not show the real pattern speed of the wave modes. Moreover, any systematic variability could be responsible for the horizontal features seen in the spectrogram.

In any case, if there are indeed several wave modes present, it is evident that they must conspire in a specific way in order to produce a spiral arm feature that is 'apparently co-rotating' as is clearly shown in Fig. \ref{omgp}. Associating the multiple wave modes to the apparent spiral arm features is beyond the scope of this paper. We rather focus on how this apparently co-rotating spiral arm feature affects the stellar motion. Therefore, we discuss the remainder of the paper in terms of what we see in Fig. \ref{omgp}. We term this 'apparent pattern speed' and 'apparent co-rotation'.




A direct implication of a decreasing apparent pattern speed is that there are co-rotation radii all over the disc, and so the radial migration (as described by \citealt{SB02}) mentioned in Section 1 is expected to occur at a wide range of radii of the spiral arm. As described in Section 1, radial migration at the co-rotation radius has been predicted to preserve the circular motion of orbits (not to heat them kinematically). 

In order to see how stellar motions are affected by this spiral arm whose apparent pattern speed deceases with radius, we trace the motions of individual particles in our simulation, as in Fig. \ref{5kpcplots}. First, we select a sample of particles around our chosen arm of interest at 5.5 kpc radius at $t=1.75$ Gyr when the arm is most prominent. The sample is selected to be near the plane of the disc ($|z|<200$pc), and has a radial thickness of 0.25 kpc and azimuthal width of $\sim$1 radian centred on highest density point of the arm (left-middle panel of Fig. \ref{5kpcplots}). We follow their motion with respect to the spiral arm around which they were selected in Fig. \ref{5kpcplots}. The middle row shows the snapshot and smoothed normalised density plots at the time they were selected. The preceding and succeeding rows show that of 60Myrs before and after respectively, and we move in a non-inertial frame of $40 \rm km s^{-1} kpc^{-1}$ from the middle step, hence only the middle row shows explicitly the actual co-ordinates in the smoothed plots. 

\begin{figure*}
\begin{center}

\subfloat{\includegraphics[scale=0.55]{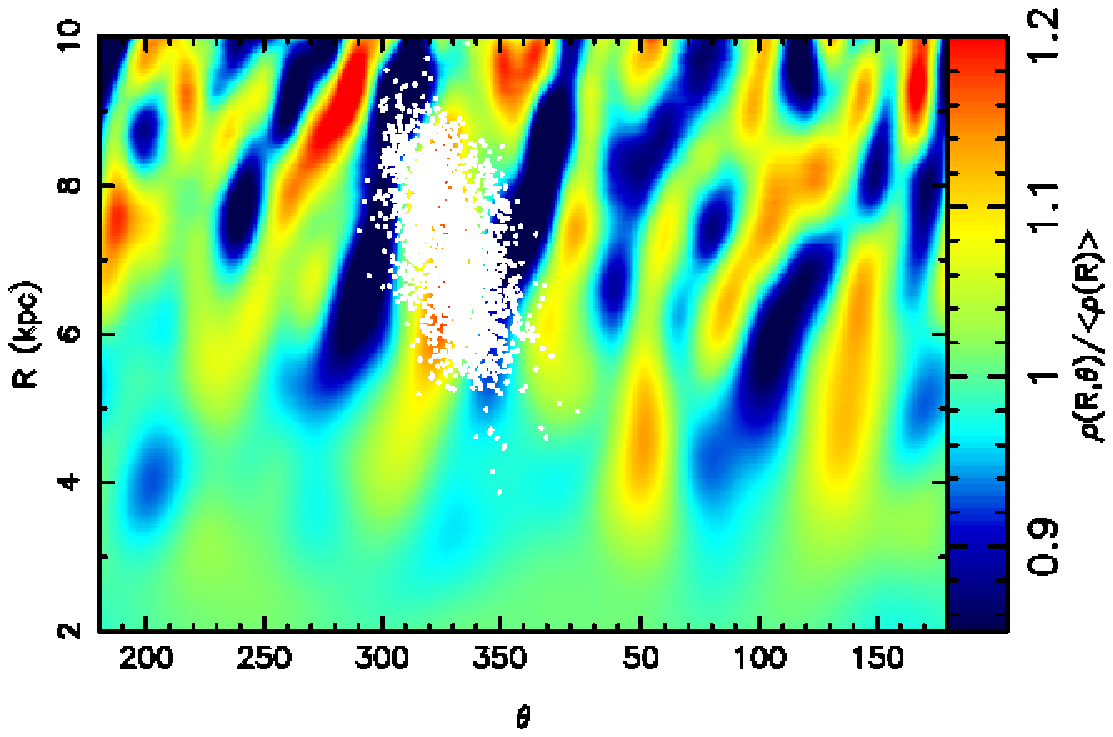}}  
\subfloat{\includegraphics[scale=0.55]{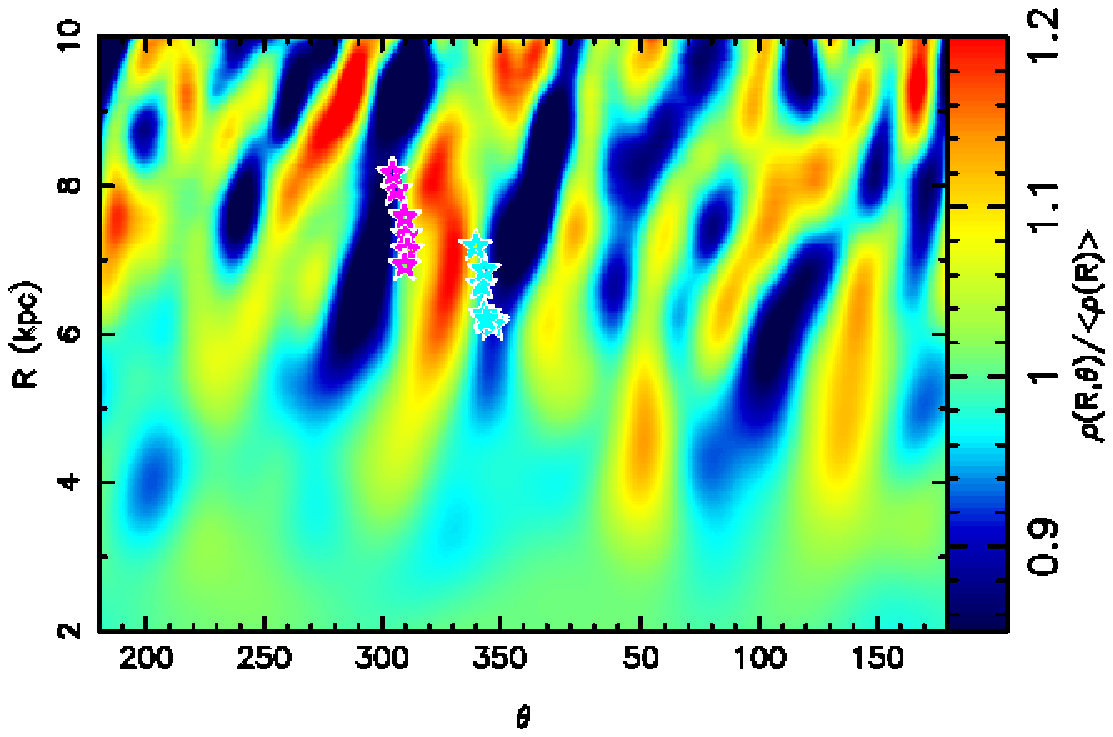}}   
 \subfloat{\includegraphics[scale=0.35]{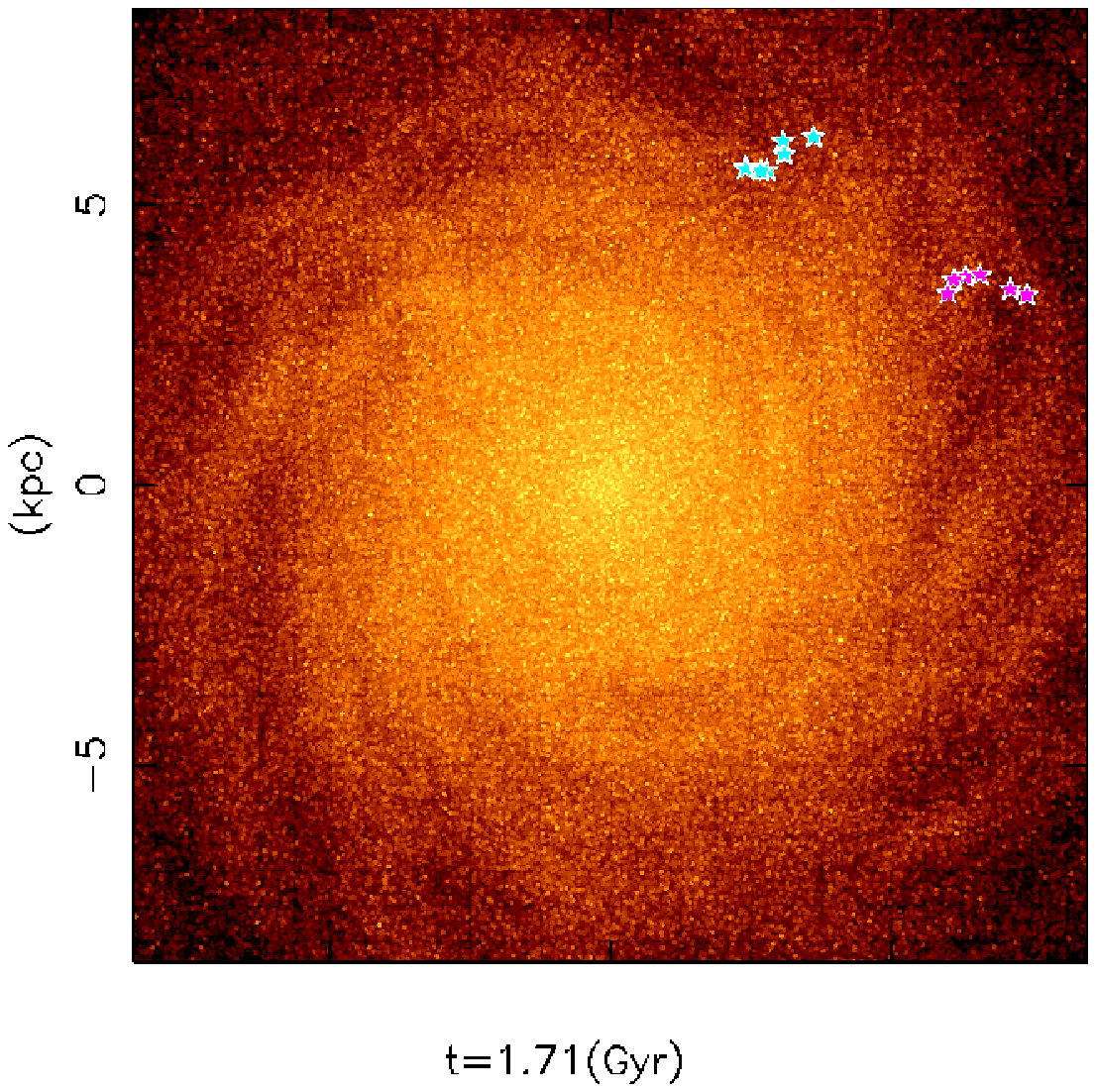}}  \\  

  \subfloat{\includegraphics[scale=0.55] {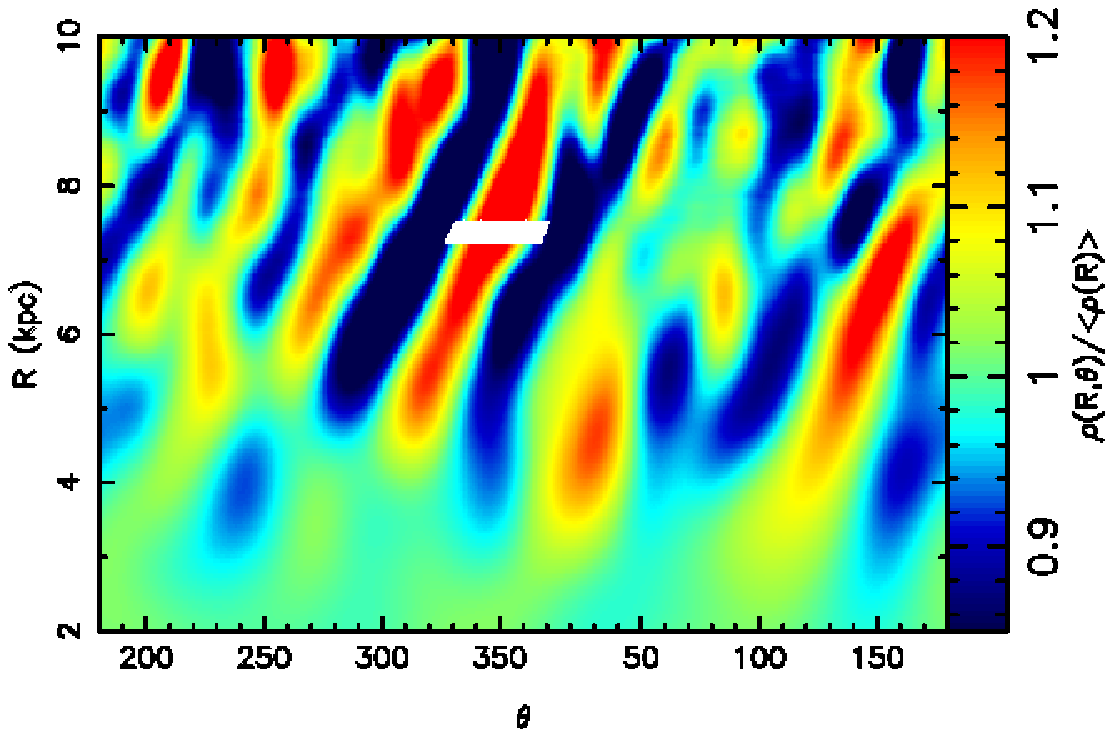}}  
   \subfloat{\includegraphics[scale=0.55] {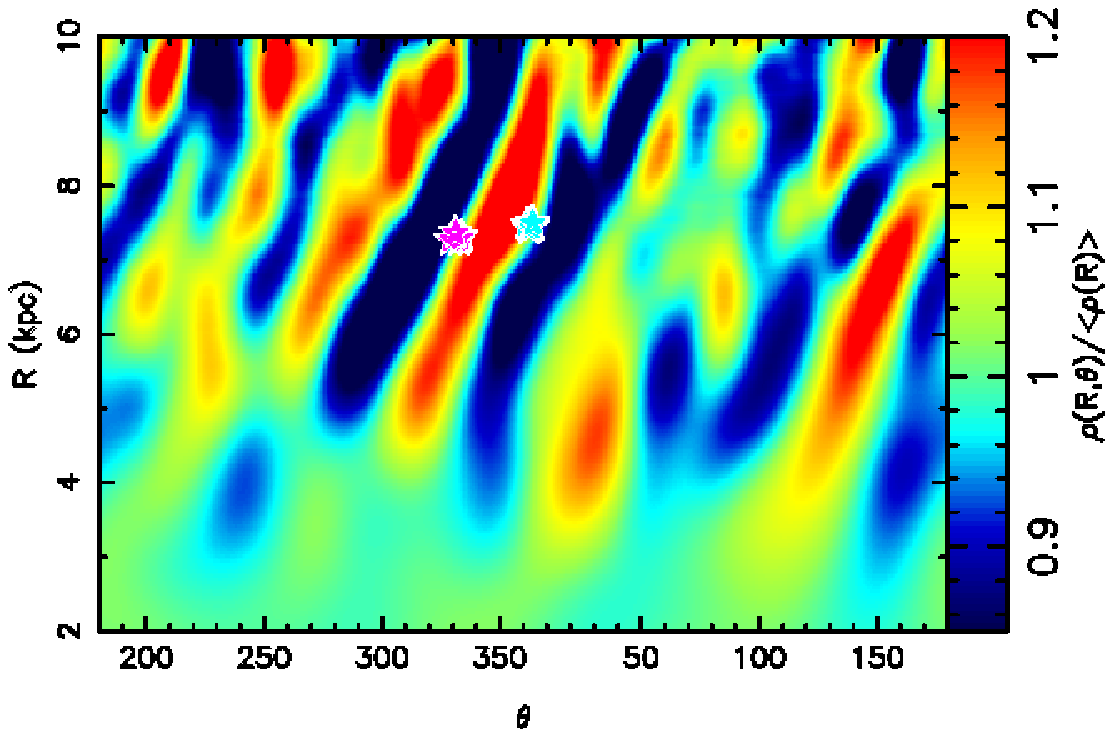}}   
    \subfloat{\includegraphics[scale=0.35]{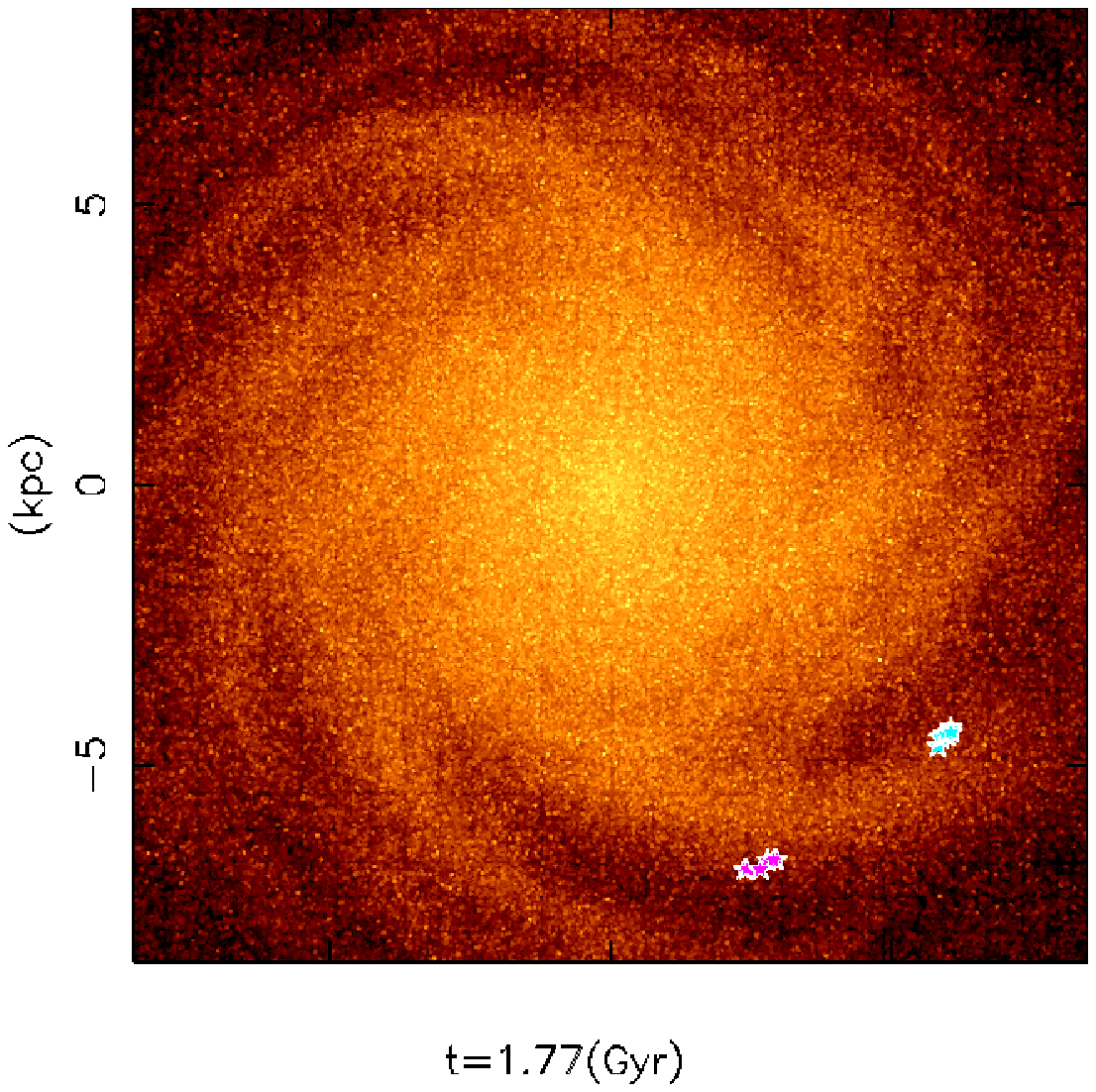}} \\      

\subfloat{\includegraphics[scale=0.55] {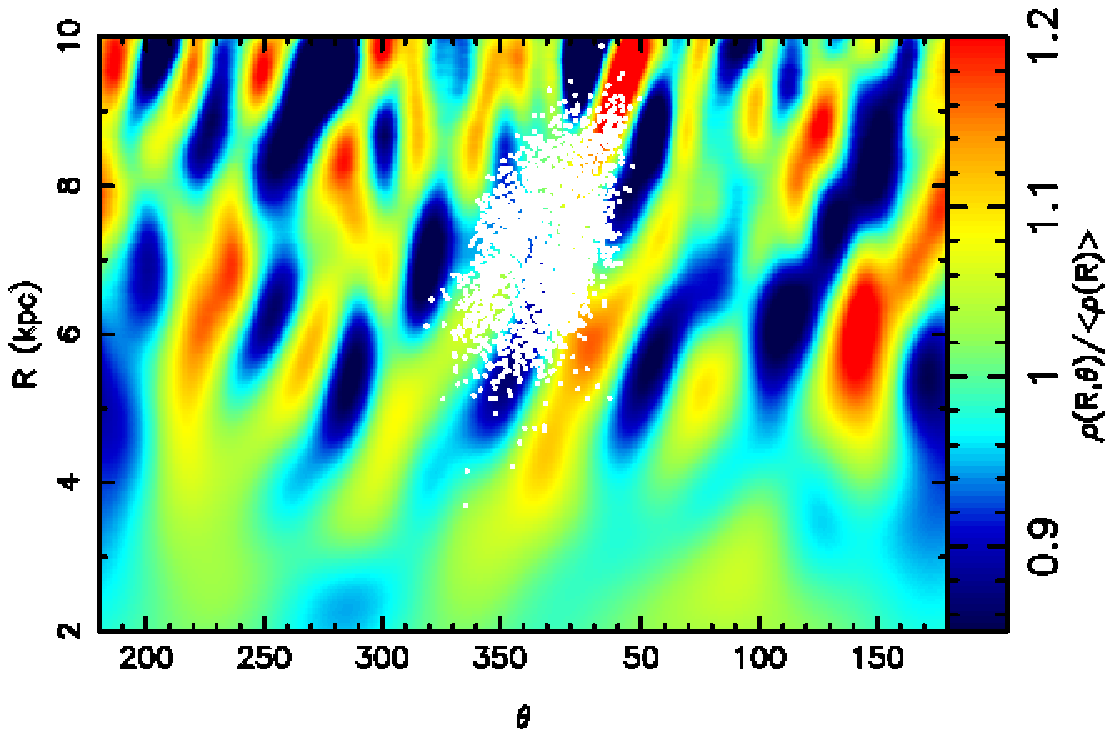}}  
\subfloat{\includegraphics[scale=0.55] {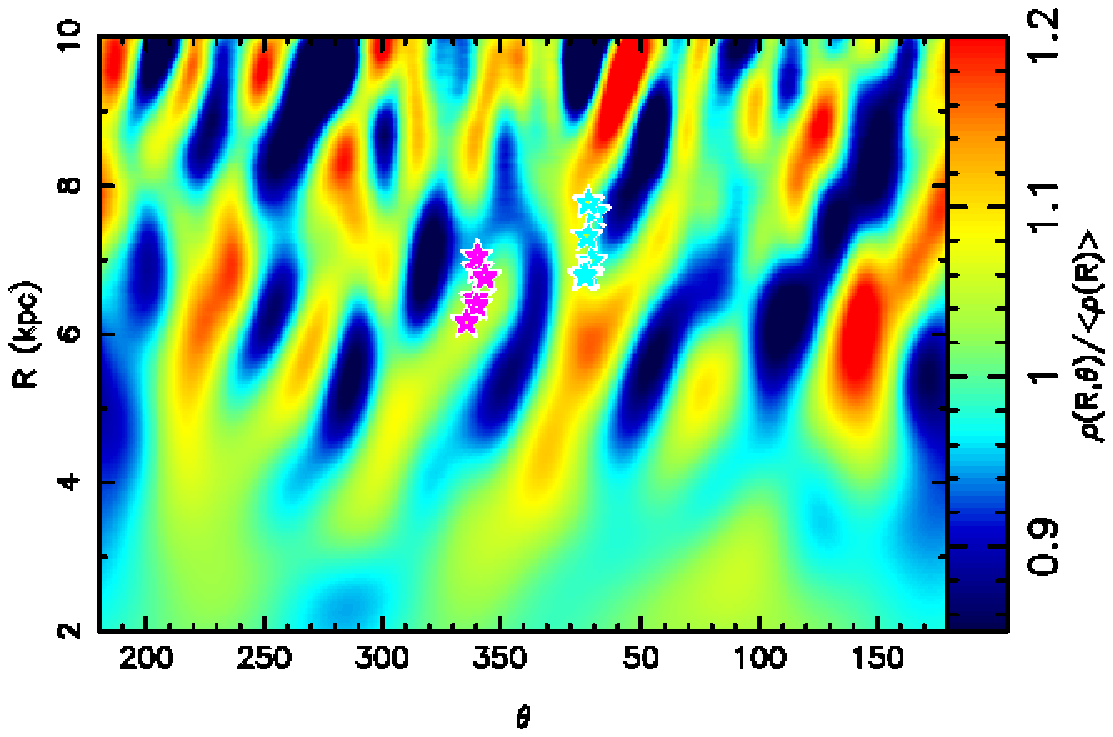}}   
\subfloat{\includegraphics[scale=0.35]{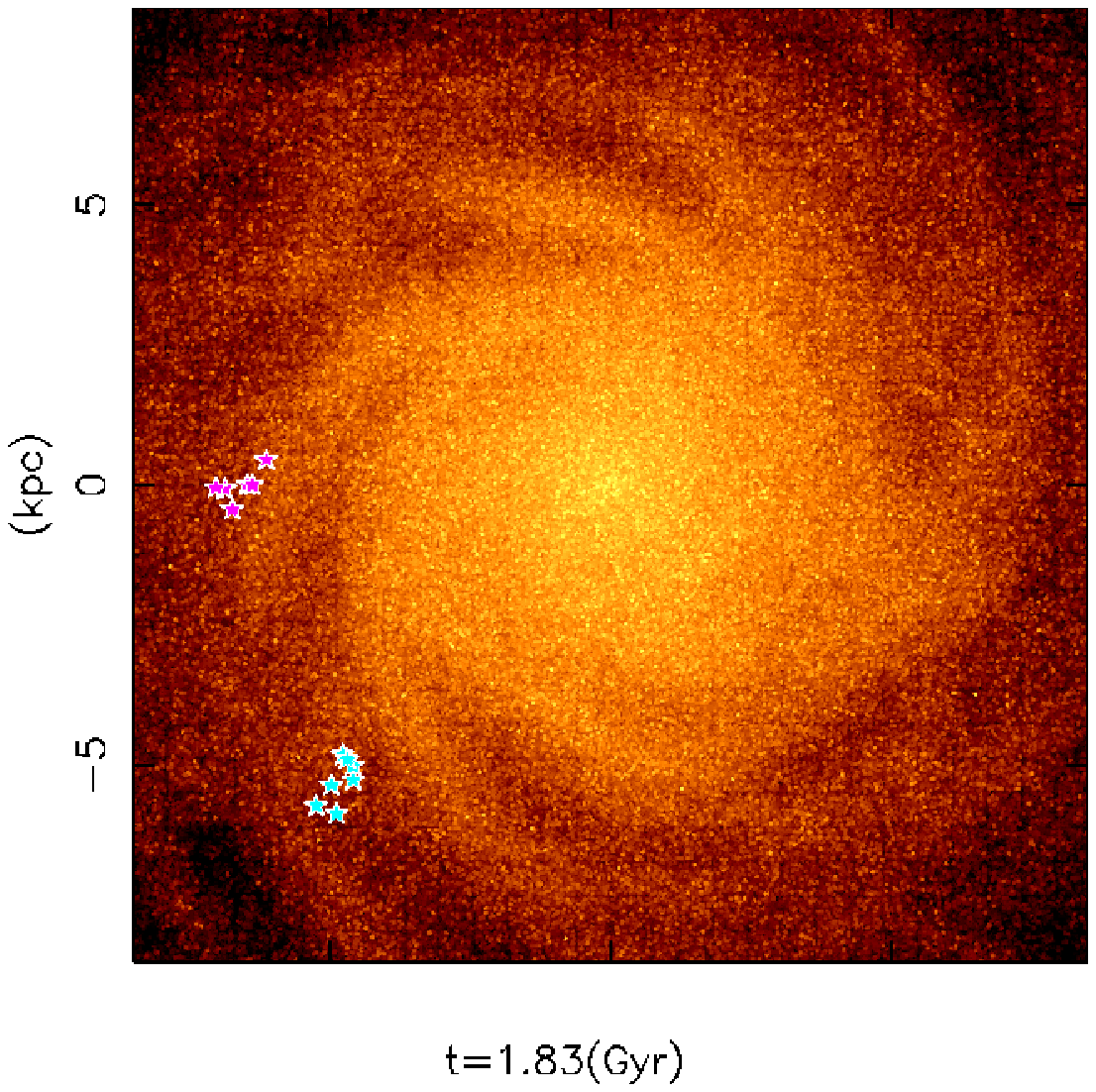}} \\      

\caption{The same as Figure \ref{5kpcplots}, but for a sample around 7.5 kpc radius. The particles were selected at 1.77 Gyr; one time step after the 5.5kpc sample, in order we see the strongest, most clear radial migration. This is because the arm at smaller radii shows more prominence at earlier times than at larger radii.}
\label{7kpcplots}
\end{center}
\end{figure*} 


It is seen that as the spiral arm grows stronger, particles from both sides of the arm begin to join the arm, which indicates the apparent co-rotation with the star particles. We find that the spiral arm develops in a way akin to swing amplification (\citealt{GLB65}; \citealt{JT66}; \citealt{T81}). In swing amplification theory, a leading arm can grow in density as it shears into a trailing arm. The star's epicycle phase and shearing motion of the arm conspire in a way that the stars in the spiral arm remain in the over-dense region for longer. This means that as the leading arm turns into a trailing arm owing to the shear motion, self-gravity becomes stronger and accumulates more stars to the arm, and the amplitude of the arm begins to grow non-linearly. 

The left column in Fig. \ref{5kpcplots} indicates that the chosen particles look like a leading feature at t=1.69 Gyr, which become part of a trailing arm at $t=1.75$ Gyr as it wound up. The selected particles around the arm at 5.5 kpc (white dots) join the arm from the outer (inner) radii at the leading (trailing) side of the spiral arm while at the same time the spiral arm appears to grow in density. We therefore witness swing amplification in action. Note however that this process is different from the classic swing amplification mechanism where the co-rotation radius is assumed to be one specific radius. What we have found may be described as swing amplification occurring over a wide range of radii. This is accompanied by strong radial migration, which we describe below.

From the selected sample of particles, we compute the angular momentum change, $\Delta L$, over a period of 80 Myrs and choose those that exhibit the largest values of $\Delta L$. As a fraction of their initial angular momentum, $L$, these have typical values $\Delta L / L \simeq 10-20\%$. We term these particles extreme migrators. The middle column of Fig. \ref{5kpcplots} shows the evolution of the extreme positive (cyan) and negative (pink) migrators. The \textquotesingle positive\textquotesingle \hspace{0.1mm} migrators are the particles that migrate towards the outer radii on the trailing side of the spiral arm. They are trapped by the potential of the spiral arm, which accelerates them. The co-rotating nature of the spiral arm feature ensures that during migration to outer radii, instead of passing through the spiral arm they remain on the trailing side (middle and right hand panels of Fig. \ref{5kpcplots}). Therefore, they continue to accelerate until the spiral arm is disrupted. The \textquotesingle negative\textquotesingle \hspace{0.1mm} migrators are particles that migrate towards the inner radii on the leading side of the spiral arm. They are decelerated as they become caught in the potential of the spiral arm, and because of the apparent co-rotation, they continue to decelerate on the leading side of the spiral arm until the spiral arm is disrupted. This illustrates the different motion that occurs on each side of the spiral arm.

To demonstrate that these stellar motions and strong migration occur over a wide range of radii, Fig. \ref{7kpcplots} shows the same dynamical evolution as Fig. \ref{5kpcplots} for a sample selected at the radius of 7.5 kpc. This 7.5 kpc sample and the extreme migrators were selected using the same criteria as the 5.5 kpc sample, however at $t=1.77$ Gyr, since we find that the spiral arm at the outer radii grows later. It is clearly demonstrated here that exactly the same type of motion expected at the co-rotation radius happens at 7.5 kpc. This motion is at least consistent with the apparent co-rotation found in Fig. \ref{omgp}, because the particles continue to join the arm from both sides at a large range of radii as they migrate, instead of passing or being passed by the spiral arm, which is expected if the pattern speed is constant as predicted by density waves. This strengthens our conclusion of apparent co-rotation of the spiral arm feature made from Fig. \ref{omgp}.

\subsection{Energy and Angular Momentum Evolution}

\begin{figure}
\centering
\includegraphics[scale=3.0]{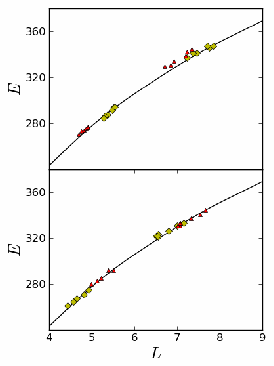}    
\caption{The energy, $E$, and angular momentum, $L$, distribution of the extreme migrators in Figs. \ref{5kpcplots} and \ref{7kpcplots} at 40 Myr before (yellow diamonds) and 40 Myr after (red triangles) the time step at which they were selected. The top (bottom) panel shows the results of the migrators that moved toward the inner (outer) radii. The solid black line indicates the circular orbit. Units are arbitrary.}
\label{extvcdblpn}
\end{figure}

When discussing the changes in angular momentum associated with radial migration, it is important to look at how the orbital energy of a star is affected. The relevant question becomes:  \textquotesingle Is the star scattered during the migration by gaining significant energy of random motion?\textquotesingle \hspace{0.1mm}. To shed light on such a question, we calculate the energy, $E$, and angular momentum, $L$, of the extreme migrators in Figs. \ref{5kpcplots} and \ref{7kpcplots} at 40 Myr before and after the time step at which they were selected, and show this in Fig. \ref{extvcdblpn}. We call these two time steps the \textquotesingle initial\textquotesingle \hspace{0.1mm} and \textquotesingle final\textquotesingle \hspace{0.1mm} time steps respectively. The solid black line indicates the $L$ and $E$ expected for a pure circular orbit at each radius. This represents the minimum energy which a star particle can have at a given angular momentum. In Fig. \ref{extvcdblpn} we show the position of the extreme migrators at the initial (yellow diamonds) and final (red triangles) time steps. We can see that the negative migrators (top panel) and positive migrators (bottom panel) move along the circular velocity curve in opposite directions to each other. Because they keep close to the circular velocity curve after migration, their orbits must retain a very similar eccentricity, and so they gain little random energy and are not scattered into higher energy orbits \citep{SB02}. 

However there is \emph{some} movement away from the circular velocity curve, and this corresponds to heating effects. We therefore test the degree of orbital heating of migrators with a larger sample of stars at different radii.  

\begin{figure}
\centering \hspace{-5.0mm}
\includegraphics[scale=1.65]{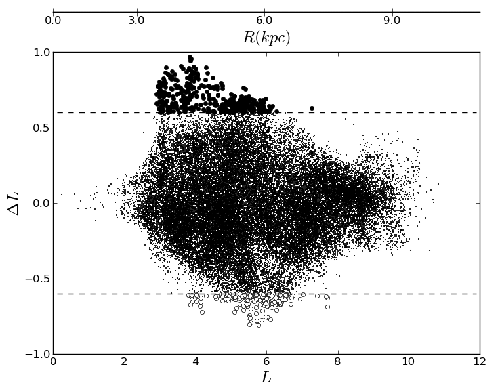}    
\caption{The angular momentum change over 80Myr for particles in a radial range of $4.0$ to $9.0$ kpc around the spiral arm at $t=1.77$ Gyr, as a function of their initial angular momentum at $t=1.73$ Gyr. We define the extreme migrators to have $|\Delta L| > 0.6$. The radius expected from circular motion with corresponding angular momentum is also shown.}
\label{dedlgt6}
\end{figure}

\begin{figure}
\centering  \hspace{-5.0mm}
\includegraphics[scale=1.68]{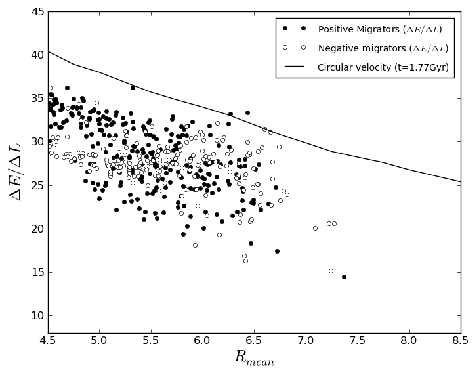}  
\caption{$\Delta E / \Delta L$ for the extreme migrators selected in Fig. \ref{dedlgt6} (open and filled circles) as a function of their mean radius between $t=1.73$ to $t=1.81$ Gyr. The filled circles are those with positive migration, and the empty are those with negative migration. The solid black line is the circular velocity. The units of y-axis or adjusted to match with that of Fig. \ref{omgp}.}
\label{Fposneg}
\end{figure}

We select the most extreme migrators from a sample of particles around the spiral arm in a radial range of 4-9 kpc at $t=1.77$ Gyr. We plot in Fig. \ref{dedlgt6} their change in angular momentum between initial and final time steps spanning a period of 80 Myr as a function of their angular momentum at the initial time step of $t=1.73$ Gyr. We selected the particles with $|\Delta L| > 0.6$, and calculate $\Delta E / \Delta L$ for each particle, which we show plotted against the mean radius of each particle over the 80 Myr period in Fig. \ref{Fposneg}.
The filled (empty) circles represent the positive (negative) migrators. A clear decreasing trend with radius is seen for both sets of migrators. This trend qualitatively matches what is expected for $\Delta E / \Delta L$ along the curve of circular orbit in Fig. \ref{extvcdblpn}, but the overall position of the particles is slightly lower. This offset is evidence of the slight \emph{heating} of the negative migrators. We note also that it is evidence for the slight \emph{cooling} for the positive migrators. This is evident in the shallower gradients of particle movement in comparison with the tangents to the circular velocity curve for both sets of migrators.

However the overall decreasing trend in parallel with the circular velocity line indicates that similar and only small heating losses occur over a range of radii. This demonstrates that radial migration can occur everywhere in the disc, and stars can slide along the spiral arms to different radii all over the disc with little contribution to disc heating.

\section{Conclusions}

We have performed N-body simulations of a pure stellar disc, and then performed a dynamical analysis of the spiral arms and particles around the spiral arms and traced their evolution. We come to the following conclusions:

\begin{enumerate}
\item{}
We find in our simulations that spiral arms are transient recurring features: we observe the continuous disappearance of spiral arms and the reappearance of new ones. This transient nature has always been found in numerical simulations, and is not consistent with spiral density wave theory.
\item{}
We performed two analyses on the pattern speed of the spiral arms: the stellar density trace method; and spectrogram analysis. The latter shows a possible configuration of two or three wave modes  that could mean that wave modes interfere constructively and destructively with each other. 

Regardless of whether this is the case, the stellar density trace method gave us clear evidence of an apparent spiral arm pattern speed that co-rotates with the stars at a large range of radii, which is reflected in points iii), iv) and v) below. This is consistent with what is found in \citet{WBS11}, and contrary to classic spiral density wave theory, which allows a constant pattern speed as a function of radius, and hence there is \emph{only one} co-rotation radius. Although we cannot reject that multiple modes are creating this apparent co-rotation, there is still the question of why the apparent pattern speed is always similar to co-rotation.
\item{}
Particles are shown to join the spiral arm from both sides at all radii. This is further evidence for the co-rotating nature of the spiral arm feature, because the arm must move at a similar speed to the particles in order for them to join the arm from both sides at all radii. 
\item{}
Particles migrate radially along the spiral arm at all radii. Stars behind the arm are accelerated by the arm and slide along the arm to larger radii. Stars in front of the arm are dragged back by the potential, and slide down the arm to smaller radii. 
\item{} 
Migrating particles do not actually cross the spiral arm. The co-rotating nature means that the particles stay on their respective sides of the spiral arm, so they are accelerated (decelerated) until the spiral arm is disrupted. This means that radial migration is more efficient than that discussed by \citet{SB02}, with little contribution to disc heating.
\end{enumerate}

Conclusions (i) and (ii) are already known from previous studies, and we include them to show the consistency of this study. Conclusions (iii), (iv) and (v) are new results which to our knowledge have not been reported before. 

A possible limitation of this study is that it focuses only on the stellar component. However, our preliminary work with simulations including gas, star formation and supernova feedback also show a consistent picture of transient co-rotating spiral arms and strong radial migration, which will be presented in a forthcoming paper. 

In this study, we have not addressed the mechanism of formation and destruction of the spiral arm structure. In this simulation, spiral arms appear to grow from small shot (Poisson) noise in the (initial) density distribution and increase in prominence from there, possibly through mechanisms similar to swing amplification. 

Although co-rotating spiral arms are expected to be short-lived due to winding, we observe that the spiral arms are disrupted before they are wound up very tightly. We also observe kinks i.e. change in pitch angle, in the spiral arms. The mechanism that controls the transient shape of the spiral arms and drives the disruption is unclear, and requires further work.

In this study, we focused on the evolution between the eras of formation and destruction, and found that the spiral arms in our simulations are not consistent with the long-lived, rigidly rotating spiral arms of spiral density wave theory. The transient co-rotating spiral arm found in N-body simulations should be tested by observations of both our Galaxy \citep{Se11} and external galaxies (e.g. \citealt{MRM09}; \citealt{FR11}; \citealt{SW11}). Although this paper focuses on the non-barred spiral galaxies, it is worthwhile to explore barred-spiral galaxies where the spiral arms are affected by the bar (e.g. \citealt{SS87}), and thus could be longer lived (e.g. \citealt{DT94} \citealt{BT08}; \citealt{BAM09}, \citealt{QDBM10}). We will study further the spiral arms in our simulations and endeavour to find observational consequences of the co-rotating spiral arm. 


\section*{acknowledgements}
The authors acknowledge the support of the UK's Science \& Technology
Facilities Council (STFC Grant ST/H00260X/1). The calculations for
this paper were performed on Cray XT4 at Center for
Computational Astrophysics, CfCA, of National Astronomical Observatory
of Japan and the DiRAC Facility jointly funded by STFC and the Large
Facilities Capital Fund of BIS. The authors acknowledge support of the
STFC funded Miracle Consortium (part of the DiRAC facility) in
providing access to the UCL Legion High Performance Computing
Facility.  The authors additionally acknowledge the support of UCL's
Research Computing team with the use of the Legion facility.
The authors thank the anonymous referee for their comments to improve the manuscript.

\bibliographystyle{mn}
\bibliography{paper01_nov14.bbl}

\begin{thebibliography}{}

\bibitem[\protect\citeauthoryear{{Amaral} \& {Lepine}}{{Amaral} \&
  {Lepine}}{1997}]{AL97}
{Amaral} L.~H.,  {Lepine} J.~R.~D.,  1997, \mnras, 286, 885

\bibitem[\protect\citeauthoryear{{Athanassoula}}{{Athanassoula}}{1984}]{A84}
{Athanassoula} E.,  1984, \physrep, 114, 319

\bibitem[\protect\citeauthoryear{{Baba}, {Asaki}, {Makino}, {Miyoshi}, {Saitoh}
  \& {Wada}}{{Baba} et~al.}{2009}]{BAM09}
{Baba} J.,  {Asaki} Y.,  {Makino} J.,  {Miyoshi} M.,  {Saitoh} T.~R.,    {Wada}
  K.,  2009, \apj, 706, 471

\bibitem[\protect\citeauthoryear{{Binney}}{{Binney}}{2010}]{Bi10}
{Binney} J.,  2010, \mnras, 401, 2318

\bibitem[\protect\citeauthoryear{{Binney} \& {Tremaine}}{{Binney} \&
  {Tremaine}}{2008}]{BT08}
{Binney} J.,  {Tremaine} S.,  2008, {Galactic Dynamics: Second Edition}.
Princeton University Press

\bibitem[\protect\citeauthoryear{{Bottema}}{{Bottema}}{2003}]{B03}
{Bottema} R.,  2003, \mnras, 344, 358

\bibitem[\protect\citeauthoryear{{Carlberg} \& {Freedman}}{{Carlberg} \&
  {Freedman}}{1985}]{CF85}
{Carlberg} R.~G.,  {Freedman} W.~L.,  1985, \apj, 298, 486

\bibitem[\protect\citeauthoryear{{Carlberg} \& {Sellwood}}{{Carlberg} \&
  {Sellwood}}{1985}]{CS85}
{Carlberg} R.~G.,  {Sellwood} J.~A.,  1985, \apj, 292, 79

\bibitem[\protect\citeauthoryear{{Dobbs} \& {Bonnell}}{{Dobbs} \&
  {Bonnell}}{2008}]{DB08}
{Dobbs} C.~L.,  {Bonnell} I.~A.,  2008, \mnras, 385, 1893

\bibitem[\protect\citeauthoryear{{Dobbs}, {Theis}, {Pringle} \& {Bate}}{{Dobbs}
  et~al.}{2010}]{DTP10}
{Dobbs} C.~L.,  {Theis} C.,  {Pringle} J.~E.,    {Bate} M.~R.,  2010, \mnras,
  403, 625

\bibitem[\protect\citeauthoryear{{Donner} \& {Thomasson}}{{Donner} \&
  {Thomasson}}{1994}]{DT94}
{Donner} K.~J.,  {Thomasson} M.,  1994, \aap, 290, 785

\bibitem[\protect\citeauthoryear{{Flynn}, {Holmberg}, {Portinari}, {Fuchs} \&
  {Jahrei{\ss}}}{{Flynn} et~al.}{2006}]{FHP06}
{Flynn} C.,  {Holmberg} J.,  {Portinari} L.,  {Fuchs} B.,    {Jahrei{\ss}} H.,
  2006, \mnras, 372, 1149

\bibitem[\protect\citeauthoryear{{Foyle}, {Rix}, {Dobbs}, {Leroy} \&
  {Walter}}{{Foyle} et~al.}{2011}]{FR11}
{Foyle} K.,  {Rix} H.-W.,  {Dobbs} C.,  {Leroy} A.,    {Walter} F.,  2011,
  ArXiv e-prints

\bibitem[\protect\citeauthoryear{{Fuchs}, {Dettbarn} \& {Tsuchiya}}{{Fuchs}
  et~al.}{2005}]{F05}
{Fuchs} B.,  {Dettbarn} C.,    {Tsuchiya} T.,  2005, \aap, 444, 1

\bibitem[\protect\citeauthoryear{{Fujii}, {Baba}, {Saitoh}, {Makino}, {Kokubo}
  \& {Wada}}{{Fujii} et~al.}{2011}]{Fu11}
{Fujii} M.~S.,  {Baba} J.,  {Saitoh} T.~R.,  {Makino} J.,  {Kokubo} E.,
  {Wada} K.,  2011, \apj, 730, 109

\bibitem[\protect\citeauthoryear{{Goldreich} \& {Lynden-Bell}}{{Goldreich} \&
  {Lynden-Bell}}{1965}]{GLB65}
{Goldreich} P.,  {Lynden-Bell} D.,  1965, \mnras, 130, 125

\bibitem[\protect\citeauthoryear{{Holmberg}, {Nordstr{\"o}m} \&
  {Andersen}}{{Holmberg} et~al.}{2009}]{HNA09}
{Holmberg} J.,  {Nordstr{\"o}m} B.,    {Andersen} J.,  2009, \aap, 501, 941

\bibitem[\protect\citeauthoryear{{Julian} \& {Toomre}}{{Julian} \&
  {Toomre}}{1966}]{JT66}
{Julian} W.~H.,  {Toomre} A.,  1966, \apj, 146, 810

\bibitem[\protect\citeauthoryear{{Kawata} \& {Gibson}}{{Kawata} \&
  {Gibson}}{2003}]{KG03}
{Kawata} D.,  {Gibson} B.~K.,  2003, \mnras, 340, 908

\bibitem[\protect\citeauthoryear{{Lin} \& {Shu}}{{Lin} \& {Shu}}{1964}]{LS64}
{Lin} C.~C.,  {Shu} F.~H.,  1964, \apj, 140, 646

\bibitem[\protect\citeauthoryear{{Lindblad}}{{Lindblad}}{1960}]{L60}
{Lindblad} P.~O.,  1960, Stockholms Observatoriums Annaler, 21, 4

\bibitem[\protect\citeauthoryear{{Lynden-Bell} \& {Kalnajs}}{{Lynden-Bell} \&
  {Kalnajs}}{1972}]{LBK72}
{Lynden-Bell} D.,  {Kalnajs} A.~J.,  1972, \mnras, 157, 1

\bibitem[\protect\citeauthoryear{{Manos} \& {Athanassoula}}{{Manos} \&
  {Athanassoula}}{2011}]{MA11}
{Manos} T.,  {Athanassoula} E.,  2011, \mnras, pp 871--+

\bibitem[\protect\citeauthoryear{{Masset} \& {Tagger}}{{Masset} \&
  {Tagger}}{1997}]{MT97}
{Masset} F.,  {Tagger} M.,  1997, \aap, 322, 442

\bibitem[\protect\citeauthoryear{{McMillan}}{{McMillan}}{2011}]{Mc11}
{McMillan} P.~J.,  2011, \mnras, 414, 2446

\bibitem[\protect\citeauthoryear{{Meidt}, {Rand} \& {Merrifield}}{{Meidt}
  et~al.}{2009}]{MRM09}
{Meidt} S.~E.,  {Rand} R.~J.,    {Merrifield} M.~R.,  2009, \apj, 702, 277

\bibitem[\protect\citeauthoryear{{Meidt}, {Rand}, {Merrifield}, {Shetty} \&
  {Vogel}}{{Meidt} et~al.}{2008}]{MRM08}
{Meidt} S.~E.,  {Rand} R.~J.,  {Merrifield} M.~R.,  {Shetty} R.,    {Vogel}
  S.~N.,  2008, \apj, 688, 224

\bibitem[\protect\citeauthoryear{{Merrifield}, {Rand} \& {Meidt}}{{Merrifield}
  et~al.}{2006}]{MRM06}
{Merrifield} M.~R.,  {Rand} R.~J.,    {Meidt} S.~E.,  2006, \mnras, 366, L17

\bibitem[\protect\citeauthoryear{{Minchev} \& {Famaey}}{{Minchev} \&
  {Famaey}}{2010}]{MF10}
{Minchev} I.,  {Famaey} B.,  2010, \apj, 722, 112

\bibitem[\protect\citeauthoryear{{Minchev}, {Famaey}, {Combes}, {Di Matteo},
  {Mouhcine} \& {Wozniak}}{{Minchev} et~al.}{2011}]{MFC11}
{Minchev} I.,  {Famaey} B.,  {Combes} F.,  {Di Matteo} P.,  {Mouhcine} M.,
  {Wozniak} H.,  2011, \aap, 527, A147+

\bibitem[\protect\citeauthoryear{{Minchev} \& {Quillen}}{{Minchev} \&
  {Quillen}}{2006}]{MQ06}
{Minchev} I.,  {Quillen} A.,  2006, in AAS/Division of Dynamical Astronomy
  Meeting \#37 Vol.~38 of Bulletin of the American Astronomical Society,
  {Radial Heating of a Galactic Disk by Multiple Spiral Density Waves}.
pp 669--+

\bibitem[\protect\citeauthoryear{{Navarro}, {Frenk} \& {White}}{{Navarro}
  et~al.}{1997}]{NFW97}
{Navarro} J.~F.,  {Frenk} C.~S.,    {White} S.~D.~M.,  1997, \apj, 490, 493

\bibitem[\protect\citeauthoryear{{Price} \& {Monaghan}}{{Price} \&
  {Monaghan}}{2007}]{PM07}
{Price} D.~J.,  {Monaghan} J.~J.,  2007, \mnras, 374, 1347

\bibitem[\protect\citeauthoryear{{Quillen}, {Dougherty}, {Bagley}, {Minchev} \&
  {Comparetta}}{{Quillen} et~al.}{2011}]{QDBM10}
{Quillen} A.~C.,  {Dougherty} J.,  {Bagley} M.~B.,  {Minchev} I.,
  {Comparetta} J.,  2011, \mnras, 417, 762

\bibitem[\protect\citeauthoryear{{Ro{\v s}kar}, {Debattista}, {Quinn},
  {Stinson} \& {Wadsley}}{{Ro{\v s}kar} et~al.}{2008}]{R08}
{Ro{\v s}kar} R.,  {Debattista} V.~P.,  {Quinn} T.~R.,  {Stinson} G.~S.,
  {Wadsley} J.,  2008, \apjl, 684, L79

\bibitem[\protect\citeauthoryear{{Ro{\v s}kar}, {Debattista}, {Stinson},
  {Quinn}, {Kaufmann} \& {Wadsley}}{{Ro{\v s}kar} et~al.}{2008}]{RDS08}
{Ro{\v s}kar} R.,  {Debattista} V.~P.,  {Stinson} G.~S.,  {Quinn} T.~R.,
  {Kaufmann} T.,    {Wadsley} J.,  2008, \apjl, 675, L65

\bibitem[\protect\citeauthoryear{{S{\'a}nchez-Bl{\'a}zquez}, {Courty}, {Gibson}
  \& {Brook}}{{S{\'a}nchez-Bl{\'a}zquez} et~al.}{2009}]{SBl09}
{S{\'a}nchez-Bl{\'a}zquez} P.,  {Courty} S.,  {Gibson} B.~K.,    {Brook} C.~B.,
   2009, \mnras, 398, 591

\bibitem[\protect\citeauthoryear{{Sellwood}}{{Sellwood}}{2010}]{Se10}
{Sellwood} J.~A.,  2010, ArXiv e-prints

\bibitem[\protect\citeauthoryear{{Sellwood}}{{Sellwood}}{2011}]{Se11}
{Sellwood} J.~A.,  2011, \mnras, 410, 1637

\bibitem[\protect\citeauthoryear{{Sellwood} \& {Binney}}{{Sellwood} \&
  {Binney}}{2002}]{SB02}
{Sellwood} J.~A.,  {Binney} J.~J.,  2002, \mnras, 336, 785

\bibitem[\protect\citeauthoryear{{Sellwood} \& {Carlberg}}{{Sellwood} \&
  {Carlberg}}{1984}]{SC84}
{Sellwood} J.~A.,  {Carlberg} R.~G.,  1984, \apj, 282, 61

\bibitem[\protect\citeauthoryear{{Sellwood} \& {Lin}}{{Sellwood} \&
  {Lin}}{1989}]{SL89}
{Sellwood} J.~A.,  {Lin} D.~N.~C.,  1989, \mnras, 240, 991

\bibitem[\protect\citeauthoryear{{Sparke} \& {Sellwood}}{{Sparke} \&
  {Sellwood}}{1987}]{SS87}
{Sparke} L.~S.,  {Sellwood} J.~A.,  1987, \mnras, 225, 653

\bibitem[\protect\citeauthoryear{{Speights} \& {Westpfahl}}{{Speights} \&
  {Westpfahl}}{2011}]{SW11}
{Speights} J.,  {Westpfahl} D.,  2011, in American Astronomical Society Meeting
  Abstracts \#218 {The Shearing HI Pattern of NGC 3031}.
pp 329.07--+

\bibitem[\protect\citeauthoryear{{Springel}, {Di Matteo} \&
  {Hernquist}}{{Springel} et~al.}{2005}]{SMH05}
{Springel} V.,  {Di Matteo} T.,    {Hernquist} L.,  2005, \mnras, 361, 776

\bibitem[\protect\citeauthoryear{{Toomre}}{{Toomre}}{1981}]{T81}
{Toomre} A.,  1981, in {S.~M.~Fall \& D.~Lynden-Bell} ed., Structure and
  Evolution of Normal Galaxies {What amplifies the spirals}.
pp 111--136

\bibitem[\protect\citeauthoryear{{Toomre}}{{Toomre}}{1990}]{T90}
{Toomre} A.,  1990, {Gas-hungry Sc spirals.}.
pp 292--303

\bibitem[\protect\citeauthoryear{{Wada}, {Baba} \& {Saitoh}}{{Wada}
  et~al.}{2011}]{WBS11}
{Wada} K.,  {Baba} J.,    {Saitoh} T.~R.,  2011, \apj, 735, 1

\end{thebibliography}

\end{document}